\documentclass[12pt,a4paper]{article}
\usepackage{amsmath,amssymb}
\usepackage[sort]{cite}
\usepackage{graphicx}
\newcommand{\ind}[2]{^{#1}_{\text{#2}}}
\def\mmu{m_{\mu}}
\def\DP{\Delta\Pi}
\def\iep{i\varepsilon}
\def\arctanh{{\rm arctanh}}
\def\arcsinh{{\rm arcsinh}}
\def\Nc{N_{{\rm c}}}
\def\nf{n_{f}}
\newcommand{\amu}[1]{a^{\text{HVP}(#1)}_{\mu}}
\newcommand{\APF}[1]{A_{0}^{(#1)}}
\newcommand{\KG}[2]{K_{#1}^{(#2)}}
\newcommand{\KGt}[2]{\tilde{K}_{#1}^{(#2)}}

\begin{document}

\begin{center}

{\Large\bf Effects of the quark flavour thresholds in \\
the hadronic vacuum polarization contributions to \\
the muon anomalous magnetic moment

}

\vskip10mm

{\large A.V.~Nesterenko}

\vskip7.5mm

{\small\it Bogoliubov Laboratory of Theoretical Physics,
Joint Institute for Nuclear Research,\\
Dubna, 141980, Russian Federation\\
E-mail:~nesterav@theor.jinr.ru}

\end{center}

\vskip5mm

\noindent
\centerline{\bf Abstract}

\vskip2.5mm

\centerline{\parbox[t]{150mm}{%
The equivalent representations for the hadronic vacuum polarization
contributions to the muon anomalous magnetic moment~$a^{\text{HVP}}_{\mu}$ in
the presence of the quark flavour thresholds are studied. Specifically, the
explicit relations between the contributions given by the integration over a
finite kinematic interval to~$a^{\text{HVP}}_{\mu}$ expressed in terms of the
hadronic vacuum polarization function, Adler function, and
the~\mbox{$R$--ratio} of electron--positron annihilation into hadrons are
derived. It is shown that the quark flavour thresholds of the hadronic vacuum
polarization function generate additional contributions
to~$a^{\text{HVP}}_{\mu}$ expressed in terms of the Adler function and
the~\mbox{$R$--ratio} and the explicit expressions for such contributions are
obtained. The~commonly employed dispersion relations, which bind together
hadronic vacuum polarization function, Adler function, and~\mbox{$R$--ratio},
are extended to account for the effects due to the quark flavour thresholds.
The~examined additional contributions due to the heavy quark thresholds
to~$a^{\text{HVP}}_{\mu}$ expressed in~terms of the~\mbox{$R$--ratio} appear
to be quite sizable, that can be of a particular relevance for the
data--driven method of assessment of the hadronic part of the muon anomalous
magnetic moment.
\\[2.5mm]
\textbf{Keywords:}~\parbox[t]{127mm}{%
muon anomalous magnetic moment, hadronic vacuum polarization contributions,
quark flavour thresholds, spacelike and timelike domains}%
}}

\vskip12mm

\section{Introduction}
\label{Sect:Intro}

The muon anomalous magnetic moment $a_{\mu} = (g_{\mu}-2)/2$ represents one
of the most intriguing issues of contemporary elementary particle physics,
which involves the whole set of interactions within the Standard Model. The
ongoing Fermilab~E989 experiment~\cite{FNAL23, FNAL21a, FNAL21b} is aiming to
achieve a four--fold increase in the measurement's accuracy with respect to
that of the precise Brookhaven E821~experiment~\cite{BNL06}, while such
planned projects as MUonE at~CERN~\cite{MUonE1, MUonE2, MUonE3, MUonE4},
E34~at~J--PARC~\cite{JPARC}, and muEDM~at~PSI~\cite{muEDM} are highly
anticipated. The~theoretical evaluations of the muon anomalous magnetic
moment (see a recent thorough review~\cite{WP20}, which is largely based on
Refs.~\cite{Davier:2010nc, Davier:2017zfy, Keshavarzi:2018mgv,
Colangelo:2018mtw, Hoferichter:2019mqg, Davier:2019can, Keshavarzi:2019abf,
Kurz:2014wya, FermilabLattice:2017wgj, Budapest-Marseille-Wuppertal:2017okr,
RBC:2018dos, Giusti:2019xct, Shintani:2019wai, FermilabLattice:2019ugu,
Gerardin:2019rua, Aubin:2019usy, Giusti:2019hkz, Melnikov:2003xd,
Masjuan:2017tvw, Colangelo:2017fiz, Hoferichter:2018kwz, Gerardin:2019vio,
Bijnens:2019ghy, Colangelo:2019uex, Pauk:2014rta, Danilkin:2016hnh,
Jegerlehner:2017gek, Knecht:2018sci, Eichmann:2019bqf, Roig:2019reh,
Colangelo:2014qya, Blum:2019ugy, Aoyama:2012wk, Aoyama:2019ryr,
Czarnecki:2002nt, Gnendiger:2013pva}) have also achieved a remarkable
precision, whereas the long--standing discrepancy of the order of a few
standard deviations between the experimental and theoretical values
of~$a_{\mu}$ may become an evidence for the existence of a new fundamental
physics beyond the Standard Model, once their accuracy is further improved.

The~uncertainty of theoretical assessment of~$a_{\mu}$ is mostly dominated by
the hadronic contribution, which involves a complex low--energy dynamics of
coloured fields, that remains unattainable within perturbation theory.
Generally, there are two methods of theoretical evaluation of the hadronic
vacuum polarization contributions to the muon anomalous magnetic
moment~$a^{\text{HVP}}_{\mu}$. In~particular, in the framework of the first
(``spacelike'') method $a^{\text{HVP}}_{\mu}$ is expressed as the integral of
either the hadronic vacuum polarization function~$\Pi(-Q^2)$ or~the related
Adler function~$D(Q^2)$ with respective kernel functions over the whole
kinematic interval~$0 \le Q^2 < \infty$. In~this case the perturbative
results for the engaged functions~$\Pi(-Q^2)$ and~$D(Q^2)$ are usually
complemented by the pertinent nonperturbative inputs, which can be
provided~by,~e.g., aforementioned MUonE direct measurements~\cite{MUonE1,
MUonE2, MUonE3, MUonE4} and lattice simulations (see~recent
reviews~\cite{Lattice1, Lattice2, Lattice3} as well as a prominent evaluation
of~$a^{\text{HVP}}_{\mu}$ reported in~Ref.~\cite{BMW21}). In~turn, in the
framework of the second (``timelike'') method~$a^{\text{HVP}}_{\mu}$ is
expressed as the integral of the \mbox{$R$--ratio} of electron--positron
annihilation into hadrons with corresponding kernel functions over the entire
kinematic interval~$s_{0} \le s < \infty$, with $s_{0}$ being the hadronic
production threshold. In~this case the perturbative results for the
function~$R(s)$ are commonly supplemented~with the low--energy experimental
data on the~\mbox{$R$--ratio}, that constitutes the data--driven method of
assessment of~$a^{\text{HVP}}_{\mu}$, see, e.g., recent
evaluations~\cite{FJ17, Keshavarzi:2019abf, Davier:2019can} (one~might also
mention here that the latest CMD--3 measurement of
$e^{+}e^{-}\to\pi^{+}\pi^{-}$ mode~\cite{CMD3}, though being in a certain
tension with its previous determinations, may eventually decrease the
discrepancy between the theoretical and experimental values of~$a_{\mu}$).

Undoubtedly, the~equivalence of the employed methods of determination of the
hadronic vacuum polarization contributions to the muon anomalous magnetic
moment is crucial for an accurate evaluation of this quantity. In~general,
the~three foregoing representations for~$a^{\text{HVP}}_{\mu}$ are commonly
derived in the assumption that the involved functions are continuous.
However, the presence of the quark flavour thresholds (that takes place, for
example, for the perturbative approximation of the functions on~hand)
violates this assumption thereby affecting the employed representations
for~$a^{\text{HVP}}_{\mu}$.

The primary objective of this paper is to derive the explicit relations
between the contributions given by the integration over a finite kinematic
interval to~$a^{\text{HVP}}_{\mu}$ expressed in terms of the
functions~$\Pi(-Q^2)$, $D(Q^2)$, and~$R(s)$, as well as to obtain the
explicit expressions for additional contributions due to the quark flavour
thresholds appearing in~$a^{\text{HVP}}_{\mu}$ expressed in terms of the
functions $D(Q^2)$ and~$R(s)$. It~is also of an apparent interest to extend
the commonly employed dispersion relations, which bind together the
functions~$\Pi(-Q^2)$, $D(Q^2)$, and~$R(s)$, to account for the effects of
the quark flavour thresholds.

The layout of the paper is as follows. Section~\ref{Sect:Methods} elucidates
the basics of the dispersion relations for the hadronic vacuum polarization
function~$\Pi(-Q^2)$, the Adler function~$D(Q^2)$, and~the function~$R(s)$,
recaps the essentials of the hadronic vacuum polarization contributions to
the muon anomalous magnetic moment, and expounds the dispersively improved
perturbation theory, which will be employed in Sect.~\ref{Sect:DPTexe} to
exemplify the results obtained in~Sect.~\ref{Sect:AmuThr}.
In~Sect.~\ref{Sect:Results} the explicit relations between the contributions
given by the integration over a finite kinematic interval
to~$a^{\text{HVP}}_{\mu}$ expressed in terms of the functions~$\Pi(-Q^2)$,
$D(Q^2)$,~$R(s)$ are obtained, the explicit expressions for additional
contributions due to the quark flavour thresholds appearing
in~$a^{\text{HVP}}_{\mu}$ expressed in terms of the functions $D(Q^2)$
and~$R(s)$ are derived, the applicability scope of the obtained results are
elucidated, the results are exemplified by making use of the dispersively
improved perturbation theory, and the commonly employed dispersion relations,
which bind together the functions~$\Pi(-Q^2)$, $D(Q^2)$,~$R(s)$, are extended
to account for the effects due to the quark flavour thresholds.
Section~\ref{Sect:Concl} summarizes the basic results.

\section{Methods}
\label{Sect:Methods}

\subsection{Basic dispersion relations}
\label{Sect:GDR}

The theoretical exploration of a variety of the strong interaction processes
is inherently based on the hadronic vacuum polarization function~$\Pi(q^2)$,
which is defined as the scalar part of the hadronic vacuum polarization
tensor
\begin{equation}
\label{P_Def}
\Pi_{\mu\nu}(q^2) = i\!\int\!d^4x\,e^{i q x} \bigl\langle 0 \bigl|\,
T\!\left\{J_{\mu}(x)\, J_{\nu}(0)\right\} \bigr| 0 \bigr\rangle =
\frac{i}{12\pi^2} (q_{\mu}q_{\nu} - g_{\mu\nu}q^2) \Pi(q^2).
\end{equation}
As argued in, e.g., Ref.~\cite{Feynman}, the
function~$\Pi(q^2)$~(\ref{P_Def}) has the only cut along the positive
semiaxis of real~$q^2$ starting at the hadronic production threshold~$q^2 \ge
s_{0}$, that enables one to write down the corresponding dispersion relation
\begin{equation}
\label{PDisp}
\DP(q^{2},q_{0}^{2}) = \Pi(q^2) - \Pi(q_0^2) =
(q^2-q_{0}^{2})\!\int\limits_{s_{0}}^{\infty}\!
\frac{R(\sigma)}{(\sigma-q^2)(\sigma-q_0^2)}\, d\sigma,
\end{equation}
with the once--subtracted Cauchy's integral formula being employed. In~this
equation $R(s)$ stands for the discontinuity of the hadronic vacuum
polarization function across the physical cut
\begin{equation}
\label{RDefP}
R(s) = \frac{1}{2 \pi i} \lim_{\varepsilon \to 0_{+}}
\Bigl[\Pi(s + \iep) - \Pi(s - \iep)\Bigr]^{\!},
\end{equation}
which is commonly identified with the so--called $R$--ratio of
electron--positron annihilation into hadrons $R(s) = \sigma(e^{+}e^{-} \to
\text{hadrons}; s)/\sigma(e^{+}e^{-} \to \mu^{+}\mu^{-}; s)$, with
\mbox{$s=q^2>0$} being the timelike kinematic variable, namely, the
center--of--mass energy squared.

In~practical applications it proves to be particularly convenient to deal
with the Adler function~\cite{Adler}
\begin{equation}
\label{GDR_DP}
D(Q^2) = -\,\frac{d\, \Pi(-Q^2)}{d \ln Q^2},
\end{equation}
where~$Q^2=-q^2>0$ denotes the spacelike kinematic variable. The pertinent
dispersion relation follows directly from Eqs.~(\ref{PDisp})
and~(\ref{GDR_DP}), namely~\cite{Adler}
\begin{equation}
\label{GDR_DR}
D(Q^2) = Q^2 \!\int\limits_{s_{0}}^{\infty}\!
\frac{R(\sigma)}{(\sigma+Q^2)^2}\, d\sigma.
\end{equation}
At~the same time, the relation inverse to Eq.~(\ref{GDR_DR}) can be obtained
by integrating Eq.~(\ref{GDR_DP}) in finite limits, that yields~\cite{Rad82,
KP82}
\begin{equation}
\label{GDR_RD}
R(s) =  \frac{1}{2 \pi i} \lim_{\varepsilon \to 0_{+}}
\int\limits_{s + \iep}^{s - \iep}
D(-\xi)\,\frac{d \xi}{\xi},
\end{equation}
where the integration contour in the complex $\xi$--plane lies in the region
of analyticity of the integrand. In~turn, the relation, which expresses the
hadronic vacuum polarization function in terms of the Adler function can also
be obtained in a similar way, specifically~\cite{Pennington77, Pennington81,
Pennington84, Pivovarov91}
\begin{equation}
\label{P_Disp2}
\DP(-Q^2\!,\, -Q_0^2) = - \int\limits_{Q_0^2}^{Q^2} D(\xi)
\frac{d \xi}{\xi},
\end{equation}
where~$Q^{2}$ and~$Q_{0}^{2}$ stand for the spacelike kinematic variable
and the subtraction point, respectively.

Basically, Eqs.~(\ref{PDisp})--(\ref{P_Disp2}) constitute the complete set of
relations, which express the functions~$\Pi(q^2)$, $D(Q^2)$, and~$R(s)$ in
terms of each other. It~is necessary to outline also that the derivation of
dispersion relations~(\ref{PDisp})--(\ref{P_Disp2}) is based only on the
kinematics of the process on~hand and involves neither model--dependent
phenomenological assumptions nor additional approximations. A~detailed
description of this topic can be found in, e.g., Chap.~1 of Ref.~\cite{Book}
and references therein.

\subsection{Hadronic vacuum polarization contributions to~$a_{\mu}$}
\label{Sect:AmuHVP}

As~noted earlier, the hadronic vacuum polarization contributions to the muon
anomalous magnetic moment~$a^{\text{HVP}}_{\mu}$ can equivalently be
represented in terms of either of the functions~$\Pi(q^2)$, $D(Q^2)$,
and~$R(s)$, which have been discussed in Sect.~\ref{Sect:GDR}. Specifically,
in the \mbox{$\ell$--th}~order in the electromagnetic coupling
\begin{subequations}
\label{Amu}
\begin{align}
\label{AmuP}
\amu{\ell} & = \APF{\ell}\!\!\int\limits_{0}^{\infty}\!
\KG{\Pi}{\ell}(Q^2) \bar\Pi(Q^2) \frac{d Q^2}{4\mmu^2} =
\APF{\ell}\!\!\int\limits_{0}^{\infty}\!
\KGt{\Pi}{\ell}(\zeta) \bar\Pi(4\zeta\mmu^2) d \zeta =
\\[1.25mm]
\label{AmuD}
& = \APF{\ell}\!\!\int\limits_{0}^{\infty} \!
\KG{D}{\ell}(Q^2) D(Q^2) \frac{d Q^2}{4\mmu^2} =
\APF{\ell}\!\!\int\limits_{0}^{\infty} \!
\KGt{D}{\ell}(\zeta) D(4\zeta\mmu^2) d \zeta = \\[1.25mm]
\label{AmuR}
& = \APF{\ell}\!\!\int\limits_{s_{0}}^{\infty} \!
\KG{R}{\ell}(s) R(s) \frac{d s}{4\mmu^2} =
\APF{\ell}\!\!\int\limits_{\chi}^{\infty} \!
\KGt{R}{\ell}(\eta) R(4\eta\mmu^2) d \eta,
\end{align}
\end{subequations}
where~$\APF{\ell}$ denotes a constant prefactor, $\bar\Pi(Q^2) = \DP(0,-Q^2)
= -\Pi(-Q^2)$ is the subtracted at zero hadronic vacuum polarization function
with $\Pi(0)=0$ being assumed, \mbox{$Q^2 = -q^2 \ge 0$} and \mbox{$s = q^2
\ge 0$} stand, respectively, for the spacelike and timelike kinematic
variables, \mbox{$\zeta = Q^2/(4\mmu^2)$} and~$\eta = s/(4\mmu^2)$ denote the
dimensionless kinematic variables, \mbox{$\chi=s_{0}/(4\mmu^2)$}, whereas
$\KG{\Pi}{\ell}(Q^2)$, $\KG{D}{\ell}(Q^2)$, and~$\KG{R}{\ell}(s)$ are the
corresponding kernel functions.

\begin{figure}[t]
\centerline{\includegraphics[height=50mm,clip]{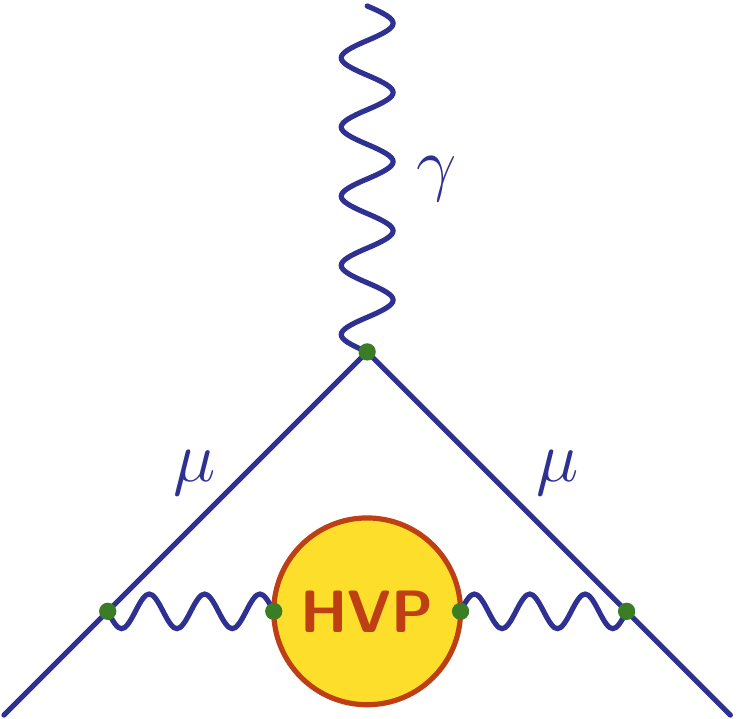}}
\caption{The leading--order hadronic vacuum polarization contribution
to the muon anomalous magnetic moment~(\ref{Amu2Def}).}
\label{Plot:Amu2}
\end{figure}

In~particular, in~the leading order of perturbation theory (i.e., in the
second order in~the electromagnetic coupling,~$\ell=2$) the corresponding
contribution is given by the diagram displayed in Fig.~\ref{Plot:Amu2}.
A~specific form of the leading--order ``timelike'' kernel
function~\cite{BKK56, BM61, KO67} makes it possible to elegantly express the
contribution~$\amu{2}$ in terms of the ``spacelike'' functions~$\bar\Pi(Q^2)$
(see Ref.~\cite{LPdR71}) and~$D(Q^2)$~(see~Refs.~\cite{Knecht2004, EdR17}) by
making use of the relevant dispersion relations, but only in this particular
case (a~discussion of this issue can be found in, e.g.,~Refs.~\cite{EdR17,
JPG49}):
\begin{align}
\label{Amu2Def}
\amu{2} & = \APF{2} \!\int\limits_{s_{0}}^{\infty}\!
G_{R}^{(2)}\!\biggl(\!\frac{s}{4\mmu^2}^{\!}\biggr)^{\!} R(s) \frac{ds}{s} =
\APF{2} \!\int\limits_{0}^{1}\! (1-x)
\bar\Pi\biggl(\!\mmu^2\frac{x^2}{1-x}\biggr) d x
\nonumber \\ & =
\APF{2} \!\int\limits_{0}^{1}\! (1-x) \biggl(\!1-\frac{x}{2}\biggr)
D\biggl(\!\mmu^2\frac{x^2}{1-x}\biggr) \frac{d x}{x},
\end{align}
where
\begin{equation}
\label{K2RInt}
\APF{2} = \frac{1}{3} \Bigl(\frac{\alpha}{\pi}\Bigr)^{\!2},
\qquad
G_{R}^{(2)}\!\biggl(\!\frac{s}{4\mmu^2}^{\!}\biggr)\! = \!
\int\limits_{0}^{1}\! \frac{x^2 (1-x)}{x^2 + (1-x) s/\mmu^2}\,dx.
\end{equation}
The ``timelike'' kernel function~$G_{R}^{(2)}(s/(4\mmu^2))$~(\ref{K2RInt})
can also be represented in explicit form~\cite{BKK56, D6263, BdR67, LdR68}
and the expression appropriate for the practical applications reads
\begin{equation}
\label{K2RExpl}
G_{R}^{(2)}(\eta) =
\eta\KGt{R}{2}(\eta) =
\frac{1}{2} + 4\eta\Bigl[(2\eta-1)\ln(4\eta)-1\Bigr]\!
-2\Bigl[2(2\eta-1)^2-1\Bigr]
\frac{A(\eta)}{\psi(\eta)},
\end{equation}
where
\begin{equation}
\label{DefAux1}
\psi(\eta) = \frac{\sqrt{\eta-1}}{\sqrt{\eta}},
\qquad
A(\eta) = \arctanh\Bigl[\psi(\eta)\Bigr],
\qquad
\eta=\frac{s}{4\mmu^2}.
\end{equation}
In~turn, the leading--order ``spacelike'' kernel
functions~$K_{\Pi}^{(2)}(Q^2)$ and~$K_{D}^{(2)}(Q^2)$ can be obtained by
mapping the integration range~$0 \le x < 1$ in~Eq.~(\ref{Amu2Def}) onto
the kinematic interval~\mbox{$0 \le Q^2 < \infty$}, namely
\begin{equation}
\label{KP2expl}
K_{\Pi}^{(2)}(4\zeta\mmu^2) = \tilde{K}_{\Pi}^{(2)}(\zeta),
\quad\,\,
G_{\Pi}^{(2)}(\zeta) =
\zeta\KGt{\Pi}{2}(\zeta) = \frac{1}{\zeta^2} \,
\frac{y^{5}(\zeta)}{1-y(\zeta)},
\quad\,\,
y(\zeta) = \zeta \Bigl(\sqrt{1+\zeta^{-1}}-1\Bigr),
\end{equation}
and
\begin{equation}
\label{KD2expl}
K_{D}^{(2)}(4\zeta\mmu^2) = \tilde{K}_{D}^{(2)}(\zeta),
\quad\,\,
G_{D}^{(2)}(\zeta) =
\zeta\KGt{D}{2}(\zeta) = (2\zeta+1)^{2}
- 2(2\zeta+1)\sqrt{\zeta(\zeta+1)} - \frac{1}{2},
\end{equation}
where $\zeta = Q^{2}/(4\mmu^2)$, see Refs.~\cite{Pivovarov2002, Blum2003,
JPG42, EdR17} and Refs.~\cite{Pivovarov2002, EdR17}, respectively. At~the
same time, it~is necessary to emphasize that this way of derivation of the
``spacelike'' kernel functions~(\ref{KP2expl}) and~(\ref{KD2expl}) from the
``timelike'' expression~(\ref{Amu2Def}) entirely relies on the particular
form of the leading--order kernel
function~$G_{R}^{(2)}(s/(4\mmu^2))$~(\ref{K2RInt}) and cannot be performed in
a general case.

\bigskip

As it has recently been proved in Ref.~\cite{JPG49}, the kernel
functions~$K_{\Pi}(Q^2)$, $K_{D}(Q^2)$, and~$K_{R}(s)$ entering
Eq.~(\ref{Amu}) can all be expressed in terms of each other. Specifically, in
an arbitrary order~$\ell$ in the electromagnetic coupling the following
relations\footnote{The relation~(\ref{KRelPR}) has also been independently
derived in a different way in~Ref.~\cite{BLP}.} hold~\cite{JPG49}:
\begin{align}
\label{KRelPR}
\KG{\Pi}{\ell}(Q^2) & = \frac{1}{2 \pi i} \lim_{\varepsilon \to 0_{+}}
\Bigl[ \KG{R}{\ell}(-Q^2+\iep) - \KG{R}{\ell}(-Q^2-\iep)\Bigr]\! =
\\[1.5mm]
\label{KRelPD}
& = - \biggl[\KG{D}{\ell}(Q^2)
+ \frac{d\,\KG{D}{\ell}(Q^2)}{d\,\ln Q^2}\biggr],
\qquad
Q^2 \ge 0,
\end{align}
\vskip-5mm
\begin{align}
\label{KRelRP}
\KG{R}{\ell}(s) & = \frac{1}{s} \int\limits_{0}^{\infty}
\KG{\Pi}{\ell}(Q^2) \frac{Q^2}{s + Q^2}\, d Q^2 =
\\[1mm]
\label{KRelRD}
& = \int\limits_{0}^{\infty}
\KG{D}{\ell}(Q^2) \frac{Q^2}{(s + Q^2)^2}\, d Q^2,
\qquad
s \ge 0,
\end{align}
\vskip-5mm
\begin{align}
\label{KRelDP}
\KG{D}{\ell}(Q^2) & = \frac{1}{Q^2}
\int\limits_{Q^2}^{\infty}\!\! \KG{\Pi}{\ell}(\xi)\, d \xi
=
\frac{4\mmu^{2}}{Q^2}K_{0}^{(\ell)} -
\frac{1}{Q^2}
\int\limits_{0}^{Q^2}\!\! \KG{\Pi}{\ell}(\xi)\, d \xi =
\\[-1mm]
\label{KRelDR}
& = - \frac{1}{2 \pi i} \lim_{\varepsilon \to 0_{+}}
\frac{1}{Q^2}
\int\limits_{Q^2 + \iep}^{Q^2 - \iep}
\KG{R}{\ell}(-p^2) d p^2.
\end{align}
In~Eq.~(\ref{KRelDP}) $\xi = -p^2 \ge 0$ stands for a spacelike kinematic
variable of the dimension of~$\mbox{GeV}^2$ and $K_{0}^{(\ell)}$~denotes
the infrared limiting value of the respective ``spacelike'' and ``timelike''
functions, namely~\cite{JPG49}
\begin{equation}
\label{KRDlim}
K_{0}^{(\ell)} =
\lim_{Q^2 \to 0_{+}} \frac{Q^2}{4\mmu^{2}}\, \KG{D}{\ell}(Q^2) =
\lim_{s \to 0_{+}} \frac{s}{4\mmu^{2}}\, \KG{R}{\ell}(s) =
\int\limits_{0}^{\infty}\!\! \KG{\Pi}{\ell}(\xi)\, \frac{d \xi}{4\mmu^{2}},
\end{equation}
whereas in Eq.~(\ref{KRelDR}) the integration contour lies in the region of
analyticity of the function~$\KG{R}{\ell}(-p^2)$, see Ref.~\cite{JPG49} for
the details. The~equations (\ref{KRelPR})--(\ref{KRelDR}) constitute the
complete set of relations, which mutually express the ``spacelike'' and
``timelike'' kernel functions~$K_{\Pi}(Q^2)$, $K_{D}(Q^2)$, and~$K_{R}(s)$
appearing in Eq.~(\ref{Amu}) in terms of each other. In~particular, it is
straightforward to verify that the foregoing kernel
functions~(\ref{K2RExpl}), (\ref{KP2expl}), and~(\ref{KD2expl}) explicitly
satisfy all six relations~(\ref{KRelPR})--(\ref{KRelDR}). It~is worthwhile to
mention also that these relations have been employed in Refs.~\cite{JPG49,
BLP} to calculate the ``spacelike'' kernel functions~$K_{\Pi}(Q^2)$
and~$K_{D}(Q^2)$ beyond the leading--order.

\subsection{Dispersively improved perturbation theory}
\label{Sect:DPT}

Factually, the dispersion relations~(\ref{PDisp})--(\ref{P_Disp2}) impose a
number of strict physical inherently nonperturbative constraints on the
functions~$\Pi(q^2)$, $R(s)$, and~$D(Q^2)$, that should definitely be taken
into account when one comes out of the limits of applicability of the QCD
perturbation theory. These nonperturbative constraints have been merged, in a
self--consistent way, with corresponding perturbative input in the framework
of dispersively improved perturbation theory~(DPT)~\cite{Book, PRD88, JPG42},
thereby enabling one to overcome some intrinsic difficulties of the
perturbative approach to~QCD and extending its applicability range towards
the infrared domain, see, in particular, Chap.~4 and Chap.~5
of~Ref.~\cite{Book} and references therein for the details. The~DPT will be
employed in Sect.~\ref{Sect:DPTexe} to illustrate the results obtained in
Sect.~\ref{Sect:AmuThr}.

The dispersively improved perturbation theory provides the following unified
integral representations for the functions on hand~\cite{Book, PRD88, JPG42}:
\begin{subequations}
\label{DPT}
\begin{align}
\label{DPT:P}
\DP(q^2,\, q_0^2) & = \Nc\!\sum_{f=1}^{\nf}\! Q_{f}^{2}\! \left[
\DP^{(0)}(q^2,\, q_0^2) +
\!\int\limits_{s_{0}}^{\infty}\!\! \rho(\sigma)
\ln\Biggl(\frac{\sigma-q^2}{\sigma-q_0^2}\,
\frac{s_{0}-q_0^2}{s_{0}-q^2}\Biggr)\frac{d\sigma}{\sigma}
\right]\!\!,
\\[1mm]
\label{DPT:R}
R(s) & = \Nc\!\sum_{f=1}^{\nf}\! Q_{f}^{2}\!\left[
R^{(0)}(s) + \theta(s-s_{0}) \!\int\limits_{s}^{\infty}\!\!
\rho(\sigma)\, \frac{d\sigma}{\sigma}
\right]\!\!,
\\[1mm]
\label{DPT:D}
D(Q^2) & = \Nc\!\sum_{f=1}^{\nf}\! Q_{f}^{2}\!\left[
D^{(0)}(Q^2) + \frac{Q^2}{Q^2+s_{0}}
\int\limits_{s_{0}}^{\infty}\!\! \rho(\sigma)\,
\frac{\sigma-s_{0}}{\sigma+Q^2}\, \frac{d\sigma}{\sigma}
\right]\!\!.
\end{align}
\end{subequations}
In~these equations $\Nc=3$ is the number of colours, $Q_{f}$~stands for the
electric charge of the quark of $f$--th flavour (in units of the elementary
charge), $\nf$~denotes the number of active flavours, $s_{0}$~is the hadronic
production threshold, $\theta(x)$~stands for the unit step--function
[$\theta(x)=1$ if $x \ge 0$ and $\theta(x)=0$ otherwise], the leading--order
terms read~\cite{Feynman, QEDAB}
\begin{subequations}
\begin{align}
\label{P0L}
\DP^{(0)}(q^2,\, q_0^2) & = 2\!\left(\!1+\frac{1}{\varkappa}\right)\!\!
\left[1-\sqrt{1+\varkappa^{-1}}\,\arcsinh\bigl(\varkappa^{1/2}\bigr)\!\right]
\nonumber \\[1mm]
& -2\!\left(\!1+\frac{1}{\varkappa_{0}}\right)\!\!
\left[1-\sqrt{1+\varkappa_{0}^{-1}}\,\arcsinh\bigl(\varkappa_{0}^{1/2}\bigr)\!\right]\!\!,
\quad
\varkappa=\frac{-q^2}{s_{0}},
\quad
\varkappa_{0}=\frac{-q_{0}^{2}}{s_{0}},
\\[1mm]
\label{R0L}
R^{(0)}(s) & = \theta(s - s_{0})\left(1-\frac{s_{0}}{s}\right)^{\!3/2}\!,
\\[1mm]
\label{D0L}
D^{(0)}(Q^2) & = 1 + \frac{3}{\xi}\!\left[1 - \sqrt{1 + \xi^{-1}}\,
\arcsinh\bigl(\xi^{1/2}\bigr)\!\right]\!\!,
\qquad \xi=\frac{Q^2}{s_{0}},
\end{align}
\end{subequations}
whereas the perturbative input is supplied by the corresponding spectral
function
\begin{equation}
\label{RhoPert}
\rho\ind{}{pert}(\sigma) = \frac{1}{\pi} \frac{d}{d\,\ln\sigma}\,
\mbox{Im}\lim_{\varepsilon \to 0_{+}} p\ind{}{pert}(\sigma-\iep) =
- \frac{d\, r\ind{}{pert}(\sigma)}{d\,\ln\sigma}
= \frac{1}{\pi}\, \mbox{Im}\lim_{\varepsilon \to 0_{+}}
d\ind{}{pert}(-\sigma-\iep),
\end{equation}
with $p\ind{}{pert}(q^2)$, $r\ind{}{pert}(s)$, and~$d\ind{}{pert}(Q^2)$ being
the perturbative expressions for the strong corrections to the
functions~$\Pi(q^2)$, $R(s)$, and~$D(Q^2)$, respectively, see
Refs.~\cite{Book, PRD88, JPG42} and references therein for the details.
In~the first order in the strong coupling the perturbative spectral
function~(\ref{RhoPert}) assumes a~quite simple form, specifically,
$\rho\ind{(1)}{pert}(\sigma) = (4/\beta_{0})
[\ln^{2}(\sigma/\Lambda^2)+\pi^2]^{-1}$, where $\beta_{0} = 11 - 2\nf/3$ is
the one--loop $\beta$~function perturbative expansion coefficient and
$\Lambda$~denotes the QCD scale parameter, whereas at the higher--loop levels
Eq.~(\ref{RhoPert}) becomes rather cumbrous (see, in particular,
Refs.~\cite{CPC1, CPC2}). Nonetheless, the explicit expression for the
perturbative spectral function~(\ref{RhoPert}) valid at an arbitrary loop
level has been obtained in Refs.~\cite{EPJC77, JPG46}, that substantially
facilitates the practical calculations within~DPT, see~Refs.~\cite{Book,
EPJC77}. It~is worthwhile to note also that the integral
representations~(\ref{DPT}) by~construction embody the aforementioned
nonperturbative constraints originated in the dispersion
relations~(\ref{PDisp})--(\ref{P_Disp2}), including the correct analytic
properties in the kinematic variables, that implies the absence of unphysical
singularities in the functions~(\ref{DPT}), see Refs.~\cite{Book, PRD88,
JPG42} and references therein for the details.

\section{Results and discussion}
\label{Sect:Results}

\subsection{Effects of the quark flavour thresholds
in~$a^{\text{\normalfont HVP}}_{\mu}$}
\label{Sect:AmuThr}

\begin{figure}[t]
\centerline{%
\includegraphics[width=75mm,clip]{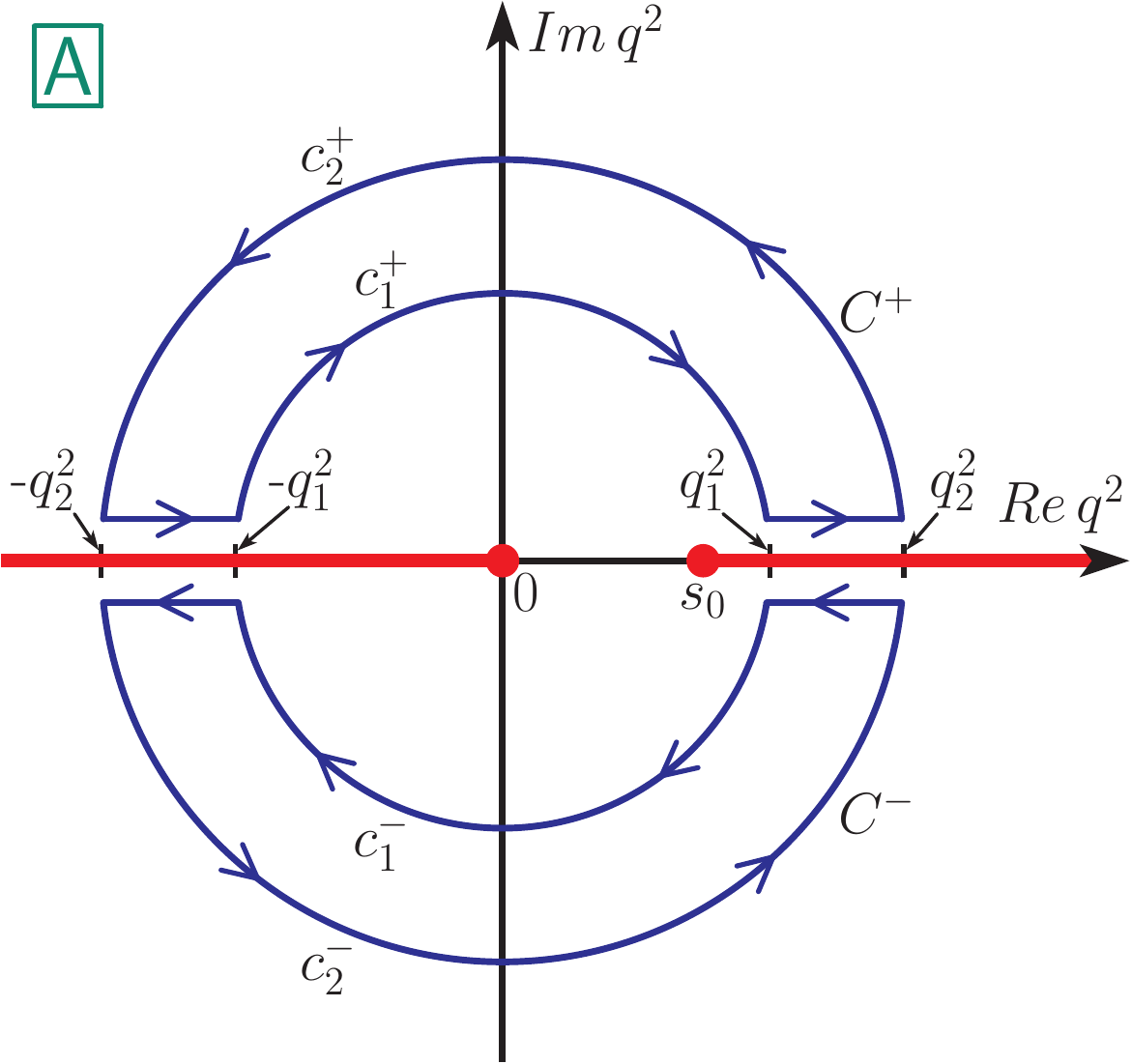}%
\hspace{10mm}%
\includegraphics[width=75mm,clip]{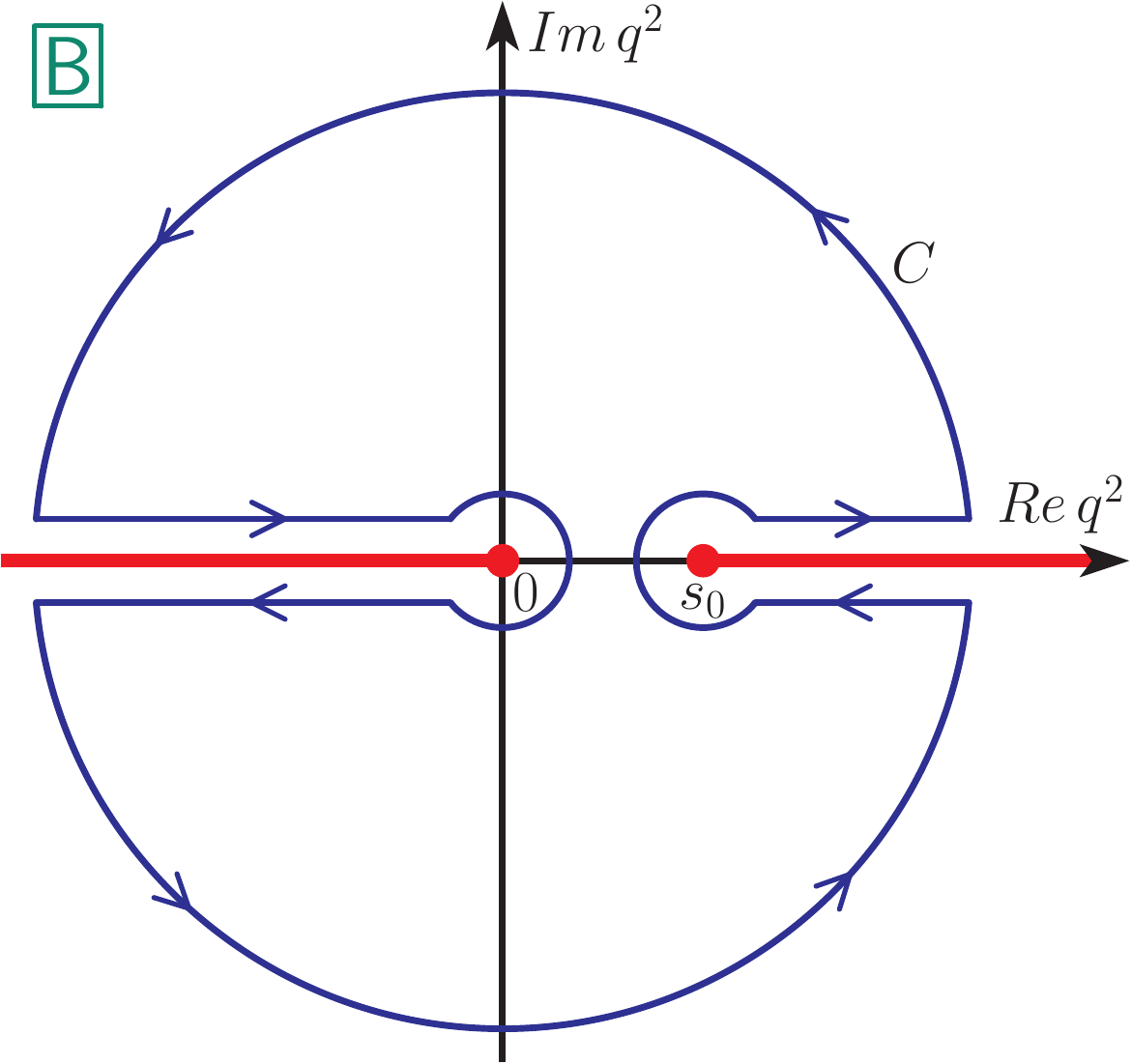}}
\caption{The closed integration contours in the complex~$q^2$--plane in
Eq.~(\ref{IntC1C2}) [contours~$C^{+}$ and~$C^{-}$, plot~``A''] and in
Eq.~(\ref{IntC}) [contour~$C$, plot~``B'']. The physical cut~$q^2 \ge s_{0}$
of the ``spacelike'' hadronic vacuum polarization
function~$\Pi(q^2)=-\bar\Pi(-q^2)$~(\ref{P_Def}) is shown along the positive
semiaxis of real~$q^2$, whereas the physical cut~\mbox{$q^2 \le 0$} of the
``timelike'' kernel function~$K_{R}(q^2)$~(\ref{AmuR}) is shown along the
negative semiaxis of real~$q^2$.}
\label{Plot:Contours}
\end{figure}

Let us first address the effects due to the quark flavour thresholds of the
hadronic vacuum polarization function in~$a^{\text{HVP}}_{\mu}$ expressed in
terms of the~\mbox{$R$--ratio} of electron--positron annihilation into
hadrons. As~noted earlier, the function~$\Pi(q^2)$~(\ref{P_Def}) has the only
cut along the positive semiaxis of real~$q^2$ starting at the hadronic
production threshold~$q^2 \ge s_{0}$, whereas the kernel function
$K_{R}(q^2)$~(\ref{AmuR}) possesses the only cut along the negative semiaxis
of real~$q^2$ starting at the origin~$q^2 \le 0$, see, e.g.,
Ref.~\cite{DispMeth5}. Thus, the integrals of the product of the
functions~$K_{R}(q^2)$~(\ref{AmuR}) and~$\Pi(q^2) = -\bar\Pi(-q^2)$ along
each of the closed contours~$C^{+}$ and~$C^{-}$ displayed in
Fig.~\ref{Plot:Contours}$\,$A vanish and, hence, so does their sum, namely
\begin{equation}
\label{IntC1C2}
\oint_{C^{+}}\! F(q^2) d q^{2} +
\oint_{C^{-}}\! F(q^2) d q^{2} = 0,
\qquad
F(q^2) = \frac{1}{4\mmu^2} K_{R}(q^2) \bar\Pi(-q^2).
\end{equation}
The~first term on the left--hand side of Eq.~(\ref{IntC1C2}) can be
represented in the following~form (the limit $\varepsilon \to 0_{+}$
is assumed hereinafter)
\begin{equation}
\label{IntC1a}
\oint_{C^{+}}\! F(q^2) d q^{2} =
\int\limits_{-q_{2}^{2}+\iep}^{-q_{1}^{2}+\iep} \! F(q^2) d q^{2} +
\int_{c_{1}^{+}}\! F(q^2) d q^{2} +
\int\limits_{q_{1}^{2}+\iep}^{q_{2}^{2}+\iep} \! F(q^2) d q^{2} +
\int_{c_{2}^{+}}\! F(q^2) d q^{2}.
\end{equation}
The change of the integration variable~$q^{2} = p^{2} + \iep$ in the first
and third terms on the right--hand side of Eq.~(\ref{IntC1a}) casts it~to
\begin{equation}
\label{IntC1b}
\oint_{C^{+}}\! F(q^2) d q^{2} =
\int\limits_{-q_{2}^{2}}^{-q_{1}^{2}} \! F(p^{2}+\iep) d p^{2} +
\int\limits_{q_{1}^{2}}^{q_{2}^{2}} \! F(p^{2}+\iep) d p^{2} +
\int_{c_{1}^{+}}\! F(q^2) d q^{2} +
\int_{c_{2}^{+}}\! F(q^2) d q^{2}.
\end{equation}
Similarly, the~second term on the left--hand side of Eq.~(\ref{IntC1C2}) can
be reduced to
\begin{equation}
\label{IntC2}
\oint_{C^{-}}\! F(q^2) d q^{2} =
\int\limits_{-q_{1}^{2}}^{-q_{2}^{2}} \! F(p^{2}-\iep) d p^{2} +
\int\limits_{q_{2}^{2}}^{q_{1}^{2}} \! F(p^{2}-\iep) d p^{2} +
\int_{c_{1}^{-}}\! F(q^2) d q^{2} +
\int_{c_{2}^{-}}\! F(q^2) d q^{2}
\end{equation}
and Eqs.~(\ref{IntC1C2}), (\ref{IntC1b}),~(\ref{IntC2}) lead~to
\begin{equation}
\label{IntC1C2b}
\int\limits_{-q_{1}^{2}}^{-q_{2}^{2}} \!
\Bigl[ F(p^{2}-\iep) - F(p^{2}+\iep) \Bigr] d p^{2} +
\int\limits_{q_{1}^{2}}^{q_{2}^{2}} \!
\Bigl[ F(p^{2}+\iep) - F(p^{2}-\iep) \Bigr]  d p^{2} +
\Delta F = 0,
\end{equation}
where
\begin{equation}
\label{DltF}
\Delta F =
\int_{c_{1}^{+}}\! F(q^2) d q^{2} +
\int_{c_{1}^{-}}\! F(q^2) d q^{2} +
\int_{c_{2}^{+}}\! F(q^2) d q^{2} +
\int_{c_{2}^{-}}\! F(q^2) d q^{2}.
\end{equation}

By~virtue of relation~(\ref{KRelPR}) the first term on the left--hand side of
Eq.~(\ref{IntC1C2b}) takes the following~form
\begin{equation}
\label{IntCSL}
\int\limits_{-q_{1}^{2}}^{-q_{2}^{2}} \!\!
\Bigl[ F(p^{2}-\iep) - F(p^{2}+\iep) \Bigr] d p^{2} =
2\pi i \! \int\limits_{Q_{1}^{2}}^{Q_{2}^{2}} \!
K_{\Pi}(Q^2) \bar\Pi(Q^2) \frac{dQ^2}{4\mmu^2},
\quad
Q_{1}^{2} = \!\left|q_{1}^{2}\right|\! \ge 0,
\quad
Q_{2}^{2} = \!\left|q_{2}^{2}\right|\! \ge 0,
\end{equation}
whereas Eq.~(\ref{RDefP}) implies that the second term on the left--hand side
of Eq.~(\ref{IntC1C2b}) can be represented~as
\begin{equation}
\label{IntCTL}
\int\limits_{q_{1}^{2}}^{q_{2}^{2}} \!
\Bigl[ F(p^{2}+\iep) - F(p^{2}-\iep) \Bigr]  d p^{2} =
- 2\pi i \! \int\limits_{s_{1}}^{s_{2}} \!
K_{R}(s) R(s) \frac{ds}{4\mmu^2},
\quad
s_{1} = q_{1}^{2} \ge 0,
\quad
s_{2} = q_{2}^{2} \ge 0,
\end{equation}
that casts Eq.~(\ref{IntC1C2b})~to
\begin{equation}
\label{RelPRint1}
\int\limits_{Q_{1}^{2}}^{Q_{2}^{2}} \!
K_{\Pi}(Q^2) \bar\Pi(Q^2) \frac{dQ^2}{4\mmu^2} =
\int\limits_{s_{1}}^{s_{2}} \!
K_{R}(s) R(s) \frac{ds}{4\mmu^2}
-
\frac{\Delta F}{2\pi i},
\end{equation}
with~$\Delta F$ being defined in Eq.~(\ref{DltF}).

Then, it is convenient to perform the integration in Eq.~(\ref{DltF}) in the
polar coordinates. For~example, the first term on the right--hand side of
Eq.~(\ref{DltF})~reads
\begin{equation}
\label{DltFc1}
\int_{c_{1}^{+}}\! F(q^2) d q^{2} = i \!\int\limits_{\pi-\varepsilon}^{\varepsilon}
\!\!H(q_{1}^{2},\varphi) d\varphi,
\qquad
H(q^{2},\varphi) =
G_{R}\!\left(\frac{q^2}{4\mmu^2}e^{i\varphi}\!\right)\!
\bar\Pi(-q^{2}e^{i\varphi}),
\end{equation}
where
\begin{equation}
\label{GRdef}
G_{R}(\eta) = \eta \tilde{K}_{R}(\eta),
\qquad
\tilde{K}_{R}(\eta) = K_{R}(4\eta\mmu^2),
\qquad
\eta = \frac{s}{4\mmu^2},
\qquad
s=q^2 \ge 0.
\end{equation}
Therefore, Eq.~(\ref{RelPRint1}) acquires the following~form
\begin{equation}
\label{RelPRint2}
\int\limits_{Q_{1}^{2}}^{Q_{2}^{2}} \!
K_{\Pi}(Q^2) \bar\Pi(Q^2) \frac{dQ^2}{4\mmu^2} =
\int\limits_{s_{1}}^{s_{2}} \!
K_{R}(s) R(s) \frac{ds}{4\mmu^2}
-\Bigl[ T(q_{2}^{2}) - T(q_{1}^{2}) \Bigr]\!,
\end{equation}
where
$Q_{1}^{2} = |q_{1}^{2}| \ge 0$,
$Q_{2}^{2} = |q_{2}^{2}| \ge 0$,
$s_{1} = q_{1}^{2} \ge 0$,
$s_{2} = q_{2}^{2} \ge 0$,
\begin{equation}
\label{Tdef}
T(q^{2}) = \frac{1}{2\pi}\!
\left[ \int\limits_{\varepsilon}^{\pi-\varepsilon}\!\!
H(q^{2},\varphi) d\varphi
+
\int\limits_{\pi+\varepsilon}^{2\pi-\varepsilon}\!\!
H(q^{2},\varphi) d\varphi \right]\!\!,
\end{equation}
and the function~$H(q^{2},\varphi)$ is defined
in~Eq.~(\ref{DltFc1}). It~is worthwhile to note here that both real and
imaginary parts of the ``timelike'' kernel
function~$G_{R}\bigl(s/(4\mmu^2)\bigr)$~(\ref{GRdef}) and the ``spacelike''
hadronic vacuum polarization function~$\bar\Pi(Q^{2})$ contribute
to~Eq.~(\ref{Tdef}).

In~the absence of the quark flavour thresholds (i.e.,~for the continuous
hadronic vacuum polarization function~$\bar\Pi(Q^2)$ in the entire energy
range~$0 \le Q^2 < \infty$) the integration contours~$C^{+}$ and~$C^{-}$
displayed in Fig.~\ref{Plot:Contours}$\,$A can be transformed into a single
integration contour~$C$ shown in Fig.~\ref{Plot:Contours}$\,$B, that casts
Eq.~(\ref{IntC1C2})~to
\begin{equation}
\label{IntC}
\oint_{C}\! F(q^2) d q^{2} = 0,
\end{equation}
with the function~$F(q^2)$ being defined in~Eq.~(\ref{IntC1C2}).
Equation~(\ref{IntC}) and contour displayed in Fig.~\ref{Plot:Contours}$\,$B
have been employed in~Ref.~\cite{JPG49} to derive the relation between the
``spacelike'' and~``timelike'' kernel functions~(\ref{KRelPR}).
In~particular, in this case the integrations along the edges of left and
right cuts lead~to the generally used Eq.~(\ref{AmuP}) and Eq.~(\ref{AmuR}),
respectively, whereas the integrations along the circles of infinitely large
and vanishing radii give no contribution to~Eq.~(\ref{IntC}).

On~the contrary, in~the presence of the quark flavour thresholds (i.e.,~for
the piecewise continuous hadronic vacuum polarization function) setting the
limits~$q_{1}^{2} \to 0$ and~$q_{2}^{2} \to \infty$ in~Eq.~(\ref{RelPRint2})
eventually results~in the additional contribution to~$a^{\text{HVP}}_{\mu}$
expressed in terms of the~\mbox{$R$--ratio}. Namely, in the
\mbox{$\ell$--th}~order in the electromagnetic coupling
[cf.~Eqs.~(\ref{AmuP}),~(\ref{AmuR})]
\begin{equation}
\label{AmuPR}
\amu{\ell}  = \APF{\ell}\!\!\int\limits_{0}^{\infty}\!
\KG{\Pi}{\ell}(Q^2) \bar\Pi(Q^2) \frac{d Q^2}{4\mmu^2} =
\APF{\ell}\!\!\int\limits_{s_{0}}^{\infty} \!
\KG{R}{\ell}(s) R(s) \frac{d s}{4\mmu^2} +
\Delta a_{\mu}^{R,(\ell)},
\end{equation}
where
\begin{equation}
\label{DltAmuR}
\Delta a_{\mu}^{R,(\ell)} = \!\sum_{f}\Delta a_{\mu,[f]}^{R,(\ell)},
\qquad
\Delta a_{\mu,[f]}^{R,(\ell)} =
\APF{\ell}\! \Bigl[
T^{(\ell)}_{[f]}\bigl(q_{f}^{2}\bigr) - T^{(\ell)}_{[f-1]}\bigl(q_{f}^{2}\bigr)
\Bigr]\!,
\end{equation}
\begin{equation}
T^{(\ell)}_{[f]}(q^{2}) = \frac{1}{2\pi}\!
\left[ \int\limits_{\varepsilon}^{\pi-\varepsilon}\!\!
H^{(\ell)}_{[f]}(q^{2},\varphi) d\varphi
+
\int\limits_{\pi+\varepsilon}^{2\pi-\varepsilon}\!\!
H^{(\ell)}_{[f]}(q^{2},\varphi) d\varphi \right]\!\!,
\end{equation}
\begin{equation}
\label{Hdef}
H^{(\ell)}_{[f]}\bigl(q^{2},\varphi\bigr) =
G_{R}^{(\ell)}\!\biggl(\frac{q^{2}}{4m_{\mu}^{2}}\,e^{i\varphi}\!\biggr)
\bar\Pi_{[f]}\bigl(-q^{2}e^{i\varphi}\bigr).
\end{equation}
In~these equations~$\APF{\ell}$ denotes a constant prefactor, the sum in
Eq.~(\ref{DltAmuR}) runs over the quark flavour thresholds,
$q_{f}^{2}$~stands for the squared threshold mass, and~$\bar\Pi_{[f]}(Q^2)$
indicates that the hadronic vacuum polarization function is calculated for
$\nf=f$ active flavours. As~it will be demonstrated in
Sect.~\ref{Sect:DPTexe}, in the leading order in the electromagnetic
coupling~(\mbox{$\ell=2$}) the additional contributions~(\ref{DltAmuR}) are
negative, so that their omission can result in an overestimation
of~$a^{\text{HVP}}_{\mu}$ within the ``timelike'' methods.

\bigskip

Let us now address the effects due to the quark flavour thresholds of the
hadronic vacuum polarization function in~$a^{\text{HVP}}_{\mu}$ expressed in
terms of the Adler function. In~particular, the integral of the product of
the ``spacelike'' kernel function~$K_{D}(Q^2)$~(\ref{AmuD}) and the Adler
function~$D(Q^2)$~(\ref{GDR_DP}) over a finite kinematic interval can be
represented~as
\begin{equation}
\label{RelPDint}
\int\limits_{Q_{1}^{2}}^{Q_{2}^{2}}
K_{D}(Q^2) D(Q^2) \frac{d Q^2}{4\mmu^2}
=
\int\limits_{Q_{1}^{2}}^{Q_{2}^{2}}
K_{\Pi}(Q^2) \bar\Pi(Q^2) \frac{d Q^2}{4\mmu^2}
+\!\left[
G_{D}\!\biggl(\frac{Q_{2}^{2}}{4m_{\mu}^{2}}\biggr)\bar\Pi(Q_{2}^{2})
- G_{D}\!\biggl(\frac{Q_{1}^{2}}{4m_{\mu}^{2}}\biggr)\bar\Pi(Q_{1}^{2})
\right]\!\!.
\end{equation}
In~this equation the integration by parts was used, the
relation~(\ref{KRelPD}) between the involved kernel functions was
employed, the definition~(\ref{GDR_DP}) was used,~and
\begin{equation}
\label{GDdef}
G_{D}(\zeta) = \zeta \tilde{K}_{D}(\zeta),
\qquad
\tilde{K}_{D}(\zeta) = K_{D}(4\zeta\mmu^2),
\qquad
\zeta = \frac{Q^2}{4\mmu^2},
\qquad
Q^2=-q^2 \ge 0.
\end{equation}
In~the absence of the quark flavour thresholds (i.e.,~for the continuous
hadronic vacuum polarization function~$\bar\Pi(Q^2)$ in the whole energy
range~$0 \le Q^2 < \infty$) the integration limits in Eq.~(\ref{RelPDint})
can safely be set to~$Q_{1}^{2} \to 0$ and~$Q_{2}^{2} \to \infty$, that
(since both terms in the squared brackets vanish in this case) casts its
left--hand and right--hand sides to the commonly employed~Eq.~(\ref{AmuD})
and~Eq.~(\ref{AmuP}), respectively. However, the presence of the quark
flavour thresholds (i.e.,~the piecewise continuous hadronic vacuum
polarization function) eventually generates in~Eq.~(\ref{RelPDint})
the~additional contribution to~$a^{\text{HVP}}_{\mu}$ expressed in terms of
the~Adler function. Specifically, in the \mbox{$\ell$--th}~order in the
electromagnetic coupling [cf.~Eqs.~(\ref{AmuP}),~(\ref{AmuD})]
\begin{equation}
\label{AmuPD}
\amu{\ell} = \APF{\ell}\!\!\int\limits_{0}^{\infty}\!
\KG{\Pi}{\ell}(Q^2) \bar\Pi(Q^2) \frac{d Q^2}{4\mmu^2} =
\APF{\ell}\!\!\int\limits_{0}^{\infty} \!
\KG{D}{\ell}(Q^2) D(Q^2) \frac{d Q^2}{4\mmu^2} +
\Delta a_{\mu}^{D^{\!},(\ell)},
\end{equation}
where
\begin{equation}
\label{DltAmuD}
\Delta a_{\mu}^{D^{\!},(\ell)} = \!\sum_{f}\Delta a_{\mu,[f]}^{D^{\!},(\ell)},
\qquad
\Delta a_{\mu,[f]}^{D^{\!},(\ell)} =
\APF{\ell} G_{D}^{(\ell)}\!\biggl(\frac{Q_{f}^{2}}{4\mmu^{2}}\biggr)\!
\biggl[\bar\Pi_{[f]}(Q_{f}^{2}) - \bar\Pi_{[f-1]}(Q_{f}^{2})\biggr]\!.
\end{equation}
In~these equations~$\APF{\ell}$ stands for a constant prefactor, the sum in
Eq.~(\ref{DltAmuD}) runs over the quark flavour thresholds,
$Q_{f}^{2}$~denotes the squared threshold mass, and~$\bar\Pi_{[f]}(Q^2)$
indicates that the hadronic vacuum polarization function is calculated for
$\nf=f$ active flavours. As~it will be shown in Sect.~\ref{Sect:DPTexe}, in
the leading order in the electromagnetic coupling~(\mbox{$\ell=2$}) the
additional contributions~(\ref{DltAmuD}) are positive, hence their omission
can lead to an underestimation of~$a^{\text{HVP}}_{\mu}$ within the
``spacelike'' methods, which involve the Adler function.

\bigskip

It is necessary to outline that the derivation of the results obtained in
this Section [namely, Eqs.~(\ref{RelPRint2}), (\ref{AmuPR}),
(\ref{RelPDint}), and~(\ref{AmuPD})] is based solely on the cut structure of
the ``spacelike'' hadronic vacuum polarization
function~$\Pi(q^2)$~(\ref{P_Def}) and the ``timelike'' kernel function
$K_{R}(q^2)$~(\ref{AmuR}) in~the complex $q^2$--plane, the
relations~(\ref{KRelPR}) and~(\ref{KRelPD}) between the involved kernel
functions, and the definitions~(\ref{RDefP}) and~(\ref{GDR_DP}), that, in
turn, makes the applicability scope of these results quite broad.
In~particular, Eqs.~(\ref{RelPRint2}) and~(\ref{RelPDint}) can be employed
within the ``window--type'' methods, which address the contribution
to~$a^{\text{HVP}}_{\mu}$ given by the integration over only a part of the
kinematic range. Additionally, Eqs.~(\ref{RelPRint2}) and~(\ref{RelPDint})
can also be employed within the methods of assessment
of~$a^{\text{HVP}}_{\mu}$ involving perturbative results,
see~Sect.~\ref{Sect:Intro}. For~example, Eq.~(\ref{RelPDint}) can be utilized
in the framework of the ``spacelike'' method, which uses the low--energy
lattice simulation data for the Adler function complemented by its
high--energy perturbative expression. In~turn, Eq.~(\ref{RelPRint2}) can be
employed within the commonly utilized ``timelike'' data--driven method, which
uses the low--energy experimental data on the~\mbox{$R$--ratio} of
electron--positron annihilation into hadrons supplemented~with the
high--energy perturbative result for the function~$R(s)$. At~the same time,
the representation of the obtained results~(\ref{RelPRint2})
and~(\ref{RelPDint}) in the form of Eqs.~(\ref{AmuPR}) and~(\ref{AmuPD}) can
be employed together with the methods, which provide the piecewise continuous
functions~$\Pi(q^2)$, $R(s)$, and~$D(Q^2)$ applicable in the entire kinematic
range, such as dispersively improved perturbation theory~\cite{Book, PRD88,
JPG42}, see~Sect.~\ref{Sect:DPTexe}.

\subsection{Exemplification of the results}
\label{Sect:DPTexe}

\begin{figure}[t]
\centerline{%
\raisebox{5.5mm}{%
\includegraphics[width=75mm,clip]{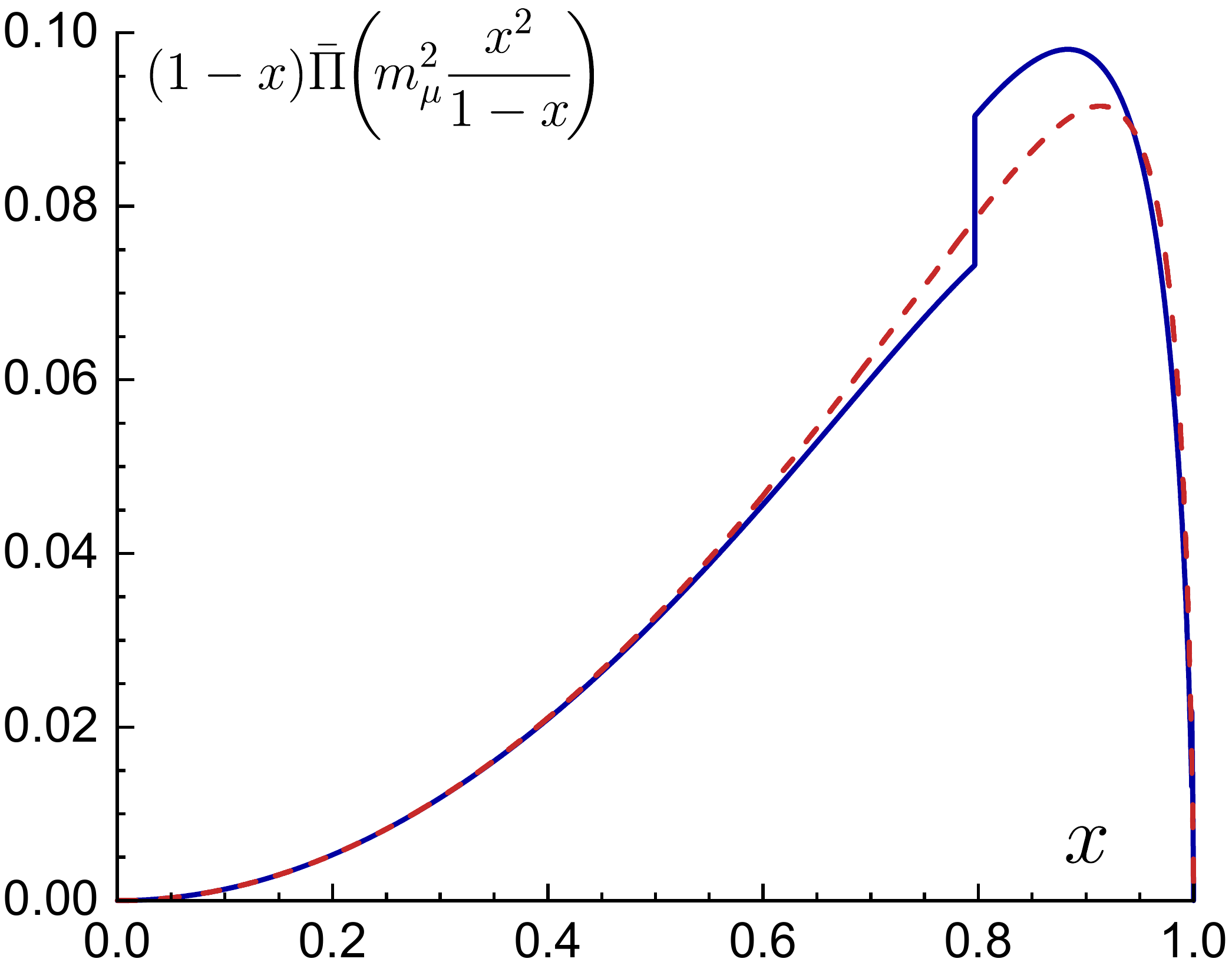}}%
\hspace{10mm}%
\includegraphics[width=75mm,clip]{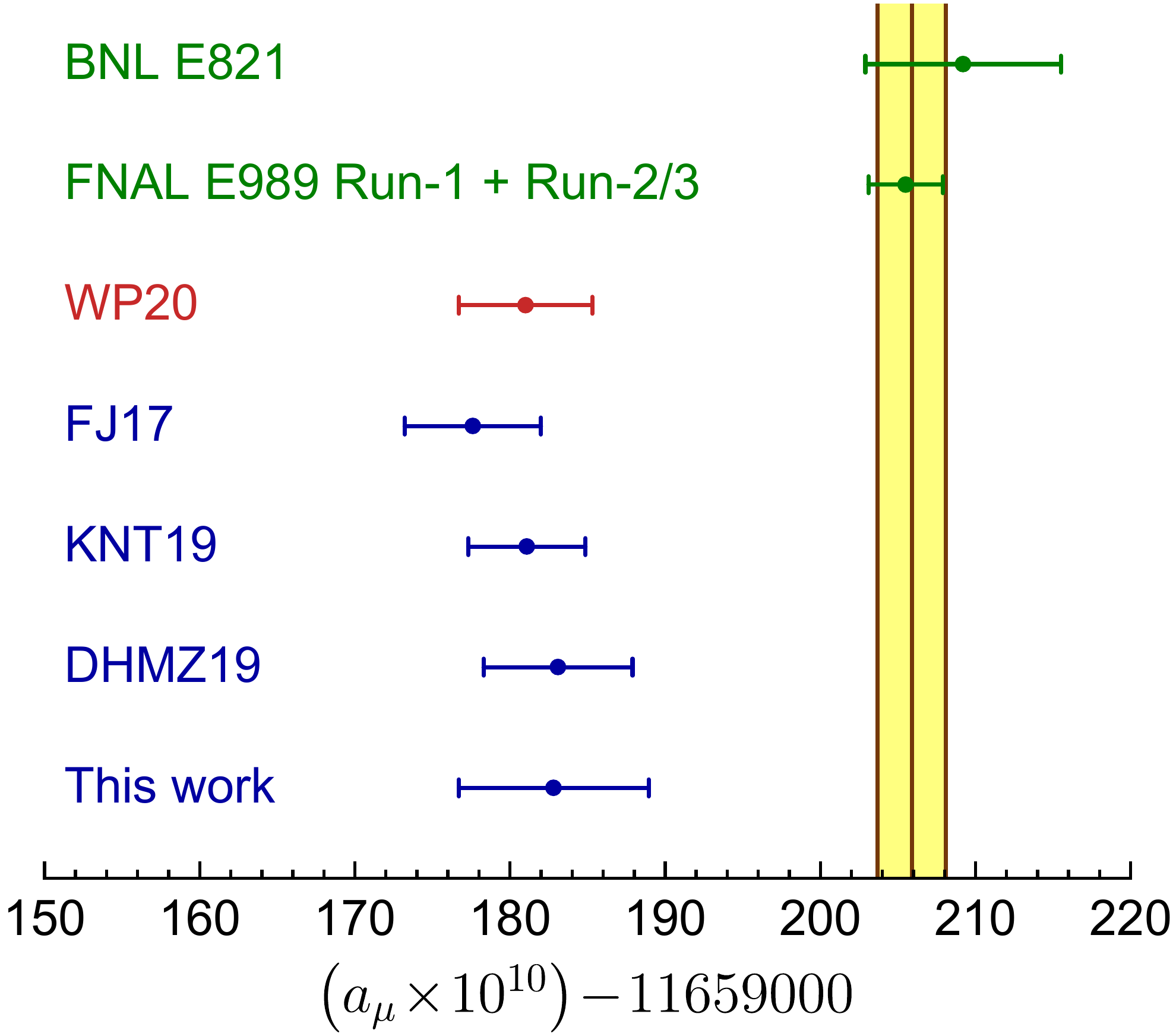}}
\caption{Left~plot: the integrands of Eq.~(\ref{Amu2Def}) corresponding to
the hadronic vacuum polarization function~(\ref{DPT:P}) (solid blue curve)
and to the data--driven one computed with the~\texttt{hadr5x19}
routine~\cite{FJ5x19} (red dashed curve). Right~plot: comparison of the
Standard Model prediction of the muon anomalous magnetic
moment~(\ref{AmuSMDPT}) with its recent assessments, namely WP20~\cite{WP20},
FJ17~\cite{FJ17}, KNT19~\cite{Keshavarzi:2019abf},
DHMZ19~\cite{Davier:2019can}. The~combined result of the
FNAL~E989~\cite{FNAL23, FNAL21a, FNAL21b} and BNL~E821~\cite{BNL06} experimental
measurements~(\ref{AmuSMexp}) is shown by vertical shaded band.}
\label{Plot:Amu}
\end{figure}

As~mentioned earlier, the dispersively improved perturbation
theory~\cite{Book, PRD88, JPG42}, which has been recapped in
Sect.~\ref{Sect:DPT}, will be~employed here to illustrate the results
obtained in~Sect.~\ref{Sect:AmuThr}. First of all, it is worthwhile to note
that the~DPT has proved to be capable of describing the hadronic vacuum
polarization contribution to the muon anomalous magnetic moment~\cite{JPG42,
QCD15, ConfXII}. Specifically, the hadronic vacuum polarization
function~(\ref{DPT:P}), being applicable in the entire energy range, can be
directly used in the representation~(\ref{AmuP}) for~$a^{\text{HVP}}_{\mu}$.
In~the leading order of perturbation theory (i.e.,~in~the second order in~the
electromagnetic coupling,~$\ell=2$) the pertinent ``spacelike'' kernel
function~$\KG{\Pi}{2}(Q^2)$ is given by Eq.~(\ref{KP2expl}), that yields
(it~is assumed that the hadronic vacuum polarization function~(\ref{DPT:P})
is in the~fourth order in the strong coupling and the
PDG22~\cite{PDG22}~values of the involved Standard Model parameters,
including the world average~$\alpha_{s}(M_{Z}^{2})$, are~used hereinafter)
\begin{equation}
\label{Amu2DPT}
\amu{2} = (695.73 \pm 5.81)\!\times\! 10^{-10},
\end{equation}
which agrees with its recent evaluations~\cite{Davier:2019can,
Keshavarzi:2019abf, FJ17, WP20}. The~left plot of Fig.~\ref{Plot:Amu}
displays the corresponding integrand of Eq.~(\ref{Amu2Def}) (solid blue
curve), whereas the red dashed curve shows the result obtained by making use
of the data--driven hadronic vacuum polarization function computed with
the~\texttt{hadr5x19} routine~\cite{FJ5x19}. The~latter represents the
function~$R(s)$ entering the dispersion relation~(\ref{PDisp}) by the
``timelike'' experimental data on the $R$--ratio of electron--positron
annihilation into hadrons at low energies and by the respective perturbative
result at high energies, that eventually smoothes out the hadronic resonances
and quark flavour thresholds. It~is necessary to outline here that, as one
can infer from the left~plot of~Fig.~\ref{Plot:Amu}, the hadronic vacuum
polarization function~(\ref{DPT:P}) can also be employed as a supplementing
infrared input for the MUonE project~\cite{MUonE1, MUonE2, MUonE3, MUonE4} in
the energy range~$x \lesssim 0.3$ uncovered by the measurements.

In~turn, in the next--to--leading order of perturbation theory (i.e.,~in~the
third order in~the electromagnetic coupling,~$\ell=3$) the required
``spacelike'' kernel functions have independently been calculated
in~Refs.~\cite{JPG49, BLP}, that results~in
\begin{equation}
\label{Amu3DPT}
\begin{split}
\amu{3a} &= (-218.02 \pm 1.35)\!\times\! 10^{-11},
\quad\,
a^{\text{HVP}(3b)}_{\mu, e} = (106.49 \pm 0.91)\!\times\! 10^{-11},
\\[1.5mm]
a^{\text{HVP}(3b)}_{\mu, \tau} &= (559.58 \pm 2.77)\!\times\! 10^{-15},
\qquad
\amu{3c} = (334.70 \pm 4.20)\!\times\! 10^{-13},
\end{split}
\end{equation}
which conform to the values reported in Refs.~\cite{Keshavarzi:2019abf, FJ17,
Kurz:2014wya, WP20}. In~Eq.~(\ref{Amu3DPT}) the $(3a)$~term embodies the
contributions of the diagrams with an additional photon line with all
possible permutations and the muon loop insertion, the $(3b)$~terms
correspond to the diagrams with an additional electron loop or $\tau$--lepton
loop insertions, whereas the~$(3c)$~term refers to the diagram with the
double hadronic insertion. The values~(\ref{Amu2DPT}) and~(\ref{Amu3DPT}),
being complemented with the remaining contributions~$a^{\text{QED}}_{\mu}$,
$a^{\text{EW}}_{\mu}$, $\amu{4}$, $a^{\text{HLbL}}_{\mu}$ given in
Ref.~\cite{WP20}, lead~to
\begin{equation}
\label{AmuSMDPT}
a^{\text{SM}}_{\mu} = (11659182.81 \pm 6.12)\!\times\! 10^{-10},
\end{equation}
which constitutes an update of the assessment performed in Refs.~\cite{JPG42,
QCD15, ConfXII}. As~one can infer from the right plot of Fig.~\ref{Plot:Amu},
the obtained result~(\ref{AmuSMDPT}) agrees with recent evaluations
of~$a^{\text{SM}}_{\mu}$~\cite{WP20, FJ17, Keshavarzi:2019abf,
Davier:2019can} and differs by $3.6$~standard deviations from the combined
result of the FNAL~E989~\cite{FNAL23, FNAL21a, FNAL21b} and BNL~E821~\cite{BNL06}
experimental measurements
\begin{equation}
\label{AmuSMexp}
a^{\text{exp}}_{\mu} = (11659205.9 \pm 2.2)\!\times\! 10^{-10}.
\end{equation}

\bigskip

Let~us turn back to the results obtained in Sect.~\ref{Sect:AmuThr}.
Factually, Eqs.~(\ref{DltAmuR}) and~(\ref{DltAmuD}) imply that in the absence
of the quark flavour thresholds (i.e.,~for a continuous hadronic vacuum
polarization function in the entire energy range) the additional
contributions~$\Delta a_{\mu}^{D^{\!},(\ell)}$ and~$\Delta
a_{\mu}^{R,(\ell)}$ vanish, that makes Eqs.~(\ref{AmuPR}) and~(\ref{AmuPD})
identical to the generally employed representations~(\ref{Amu}). Indeed, in
a~hypothetical case when the number of active flavours is~assumed to be fixed
in the whole energy range, the representations~(\ref{DPT}) provide continuous
functions~$\bar\Pi(Q^2)$, $D(Q^2)$,~$R(s)$, and their use in~either of the
expressions~(\ref{Amu}) yield the same result. For~example, in~the leading
order of perturbation theory (i.e., in the second order in~the
electromagnetic coupling,~\mbox{$\ell=2$}) for~the case of two~active
flavours~($\nf=2$) each of the representations~(\ref{Amu}) with
functions~(\ref{DPT}) leads~to (the mean values will be addressed herein)
\begin{equation}
a^{\text{HVP}(2)}_{\mu,[\nf=2]} =
a^{\text{HVP}(2)}_{\mu,\Pi} =
a^{\text{HVP}(2)}_{\mu,D} =
a^{\text{HVP}(2)}_{\mu,R} =
636.06 \!\times\! 10^{-10}.
\end{equation}
In~this equation the additional subscripts ``$\Pi$'', ``$D$'', and~``$R$''
indicate that the evaluated contribution to the muon anomalous magnetic
moment is represented in terms of the hadronic vacuum polarization
function~(\ref{AmuP}), the Adler function~(\ref{AmuD}), and the $R$--ratio of
electron--positron annihilation into hadrons~(\ref{AmuR}), respectively.
On~the contrary, in the presence of the quark flavour thresholds the
functions~$\bar\Pi(Q^2)$, $D(Q^2)$,~$R(s)$~(\ref{DPT}), likewise their
perturbative counterparts, become piecewise continuous and the
expressions~(\ref{Amu}), since the magnitudes of their integrands are
distributed differently over the corresponding kinematic ranges, yield
diverse results. Specifically, whereas the representation~(\ref{AmuP})
leads~to an earlier quoted value~(\ref{Amu2DPT}), the
expressions~(\ref{AmuD}) and~(\ref{AmuR}) result~in
\begin{equation}
\label{Amu2var}
a^{\text{HVP}(2)}_{\mu,\Pi} = 695.73 \!\times\! 10^{-10},
\qquad
a^{\text{HVP}(2)}_{\mu,D}   = 664.19 \!\times\! 10^{-10},
\qquad
a^{\text{HVP}(2)}_{\mu,R}   = 800.03 \!\times\! 10^{-10}.
\end{equation}
At~the same time, in the presence of the quark flavour thresholds the
additional contributions~(\ref{DltAmuR}) and~(\ref{DltAmuD}) become
nonvanishing
\begin{equation}
\Delta a_{\mu}^{D^{\!},(2)} =   31.54 \!\times\! 10^{-10},
\qquad
\Delta a_{\mu}^{R,(2)}      = -104.30 \!\times\! 10^{-10}
\end{equation}
and either of~Eqs.~(\ref{AmuPR}) and~(\ref{AmuPD}) yields the same
result~(\ref{Amu2DPT}), namely
\begin{equation}
\amu{2} = a^{\text{HVP}(2)}_{\mu,\Pi} =
a^{\text{HVP}(2)}_{\mu,D} + \Delta a_{\mu}^{D^{\!},(2)} =
a^{\text{HVP}(2)}_{\mu,R} + \Delta a_{\mu}^{R,(2)} =
695.73 \!\times\! 10^{-10}.
\end{equation}
It~is worthwhile to note here that, since at high energies the kernel
functions~$K_{D}(Q^2)$ and~$K_{R}(s)$ decrease as their arguments increase,
the~individual contributions~$\Delta
a_{\mu,[f]}^{D^{\!},(\ell)}$~(\ref{DltAmuD}) and~$\Delta
a_{\mu,[f]}^{R,(\ell)}$~(\ref{DltAmuR}) from the heavy quark thresholds also
become smaller as~$f$ increases, in~particular
\begin{subequations}
\begin{align}
\label{DltPDHQT}
\Delta a_{\mu,[4]}^{D^{\!},(2)} & = (135.26 \pm  7.37) \!\times\! 10^{-13},
\qquad\;\;\,
\Delta a_{\mu,[5]}^{D^{\!},(2)}   = (456.33 \pm 10.13) \!\times\! 10^{-16},
\\[1.5mm]
\label{DltPRHQT}
\Delta a_{\mu,[4]}^{R^{\!},(2)} & = (-143.97 \pm 4.46) \!\times\! 10^{-11},
\qquad
\Delta a_{\mu,[5]}^{R^{\!},(2)}   = (-337.63 \pm 4.05) \!\times\! 10^{-13},
\end{align}
\end{subequations}
whereas the effects due to the top quark are essentially negligible in the
context of the hadronic contributions to the muon anomalous magnetic moment.
It~is necessary to emphasize that for~$a^{\text{HVP}}_{\mu}$ expressed in
terms of the~\mbox{$R$--ratio}~(\ref{AmuPR}) the individual contributions
from the heavy quark thresholds~(\ref{DltPRHQT}) appear to be quite sizable,
that can be of a particular relevance for the data--driven approach.
Specifically, the contribution to~$a^{\text{HVP}}_{\mu}$~(\ref{AmuPR})
due to even the bottom quark threshold~$\Delta
a_{\mu,[5]}^{R^{\!},(2)}$~(\ref{DltPRHQT}) is as large as the discussed
earlier next--to--leading order contribution~$\amu{3c}$~(\ref{Amu3DPT}).

\subsection{Effects of the quark flavour thresholds in~dispersion relations}
\label{Sect:GDRext}

Basically, the~foregoing dispersion relations~(\ref{PDisp})--(\ref{P_Disp2}),
which express hadronic vacuum polarization function~$\Pi(q^2)$, Adler
function~$D(Q^2)$, and $R$--ratio of electron--positron annihilation into
hadrons~$R(s)$ in terms of each other, are commonly derived in the assumption
of continuity of the functions on~hand. Therefore, it is of an apparent
interest to address the relations~(\ref{PDisp})--(\ref{P_Disp2}) in the
presence of the quark flavour thresholds, i.e., for the piecewise continuous
functions~$\Pi(q^2)$, $D(Q^2)$, and~$R(s)$.

\begin{figure}[t]
\centerline{\includegraphics[width=75mm,clip]{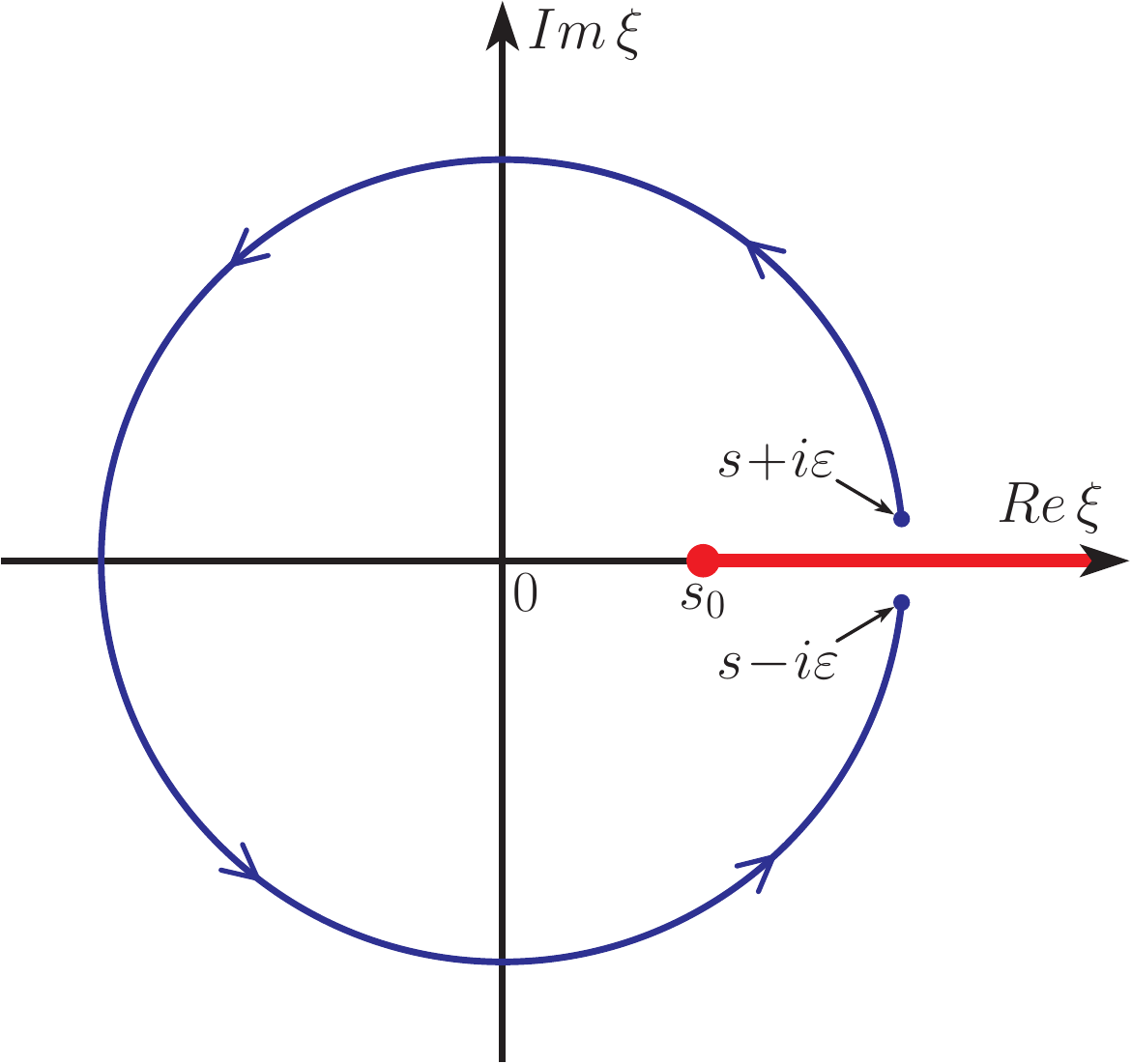}}
\caption{The integration contour in the complex $\xi$--plane in
Eq.~(\ref{GDRextRD}). The physical cut $\xi \ge s_{0}$ of the
Adler function $D(-\xi)$~(\ref{GDRextDP}) is shown along the
positive semiaxis of real~$\xi$.}
\label{Plot:CGDRextRD}
\end{figure}

First of all, the definitions~(\ref{RDefP}) and~(\ref{GDR_DP}) remain the
same, specifically
\begin{equation}
\label{GDRextRP}
R(s) = \frac{1}{2 \pi i} \lim_{\varepsilon \to 0_{+}}
\Bigl[\Pi(s + \iep) - \Pi(s - \iep)\Bigr]^{\!},
\qquad
s=q^2>0
\end{equation}
and
\begin{equation}
\label{GDRextDP}
D(Q^2) = -\,\frac{d\, \Pi(-Q^2)}{d \ln Q^2},
\qquad
Q^2=-q^2>0.
\end{equation}
Then, the relation~(\ref{GDR_RD}) also keeps its original form, though here
it becomes preferable to choose as the integration contour in the complex
$\xi$--plane the one displayed in Fig.~\ref{Plot:CGDRextRD}, that leads~to
\begin{equation}
\label{GDRextRD}
R(s) =  \frac{1}{2 \pi i} \lim_{\varepsilon \to 0_{+}}
\int\limits_{s + \iep}^{s - \iep} D(-\xi)\,\frac{d \xi}{\xi} =
\frac{1}{2\pi} \lim_{\varepsilon \to 0_{+}}
\!\int\limits_{\varepsilon}^{2\pi-\varepsilon}\!
D(-s e^{i\varphi}) d\varphi,
\end{equation}
whereas the relation~(\ref{P_Disp2}) acquires the form
\begin{equation}
\label{GDRextPD}
\DP(-Q^2\!,\, -Q_0^2) = - \int\limits_{Q_0^2}^{Q^2} D(\xi)
\frac{d \xi}{\xi} +
\sum\limits_{f} \Bigl[ \Pi_{[f]}(-Q_{f}^{2}) - \Pi_{[f-1]}(-Q_{f}^{2}) \Bigr],
\end{equation}
where the sum runs over the quark flavour thresholds located within the
integration range and the subscript~``$[f]$'' indicates that the hadronic
vacuum polarization function is calculated for $\nf=f$ active flavours.

\begin{figure}[t]
\centerline{%
\includegraphics[width=75mm,clip]{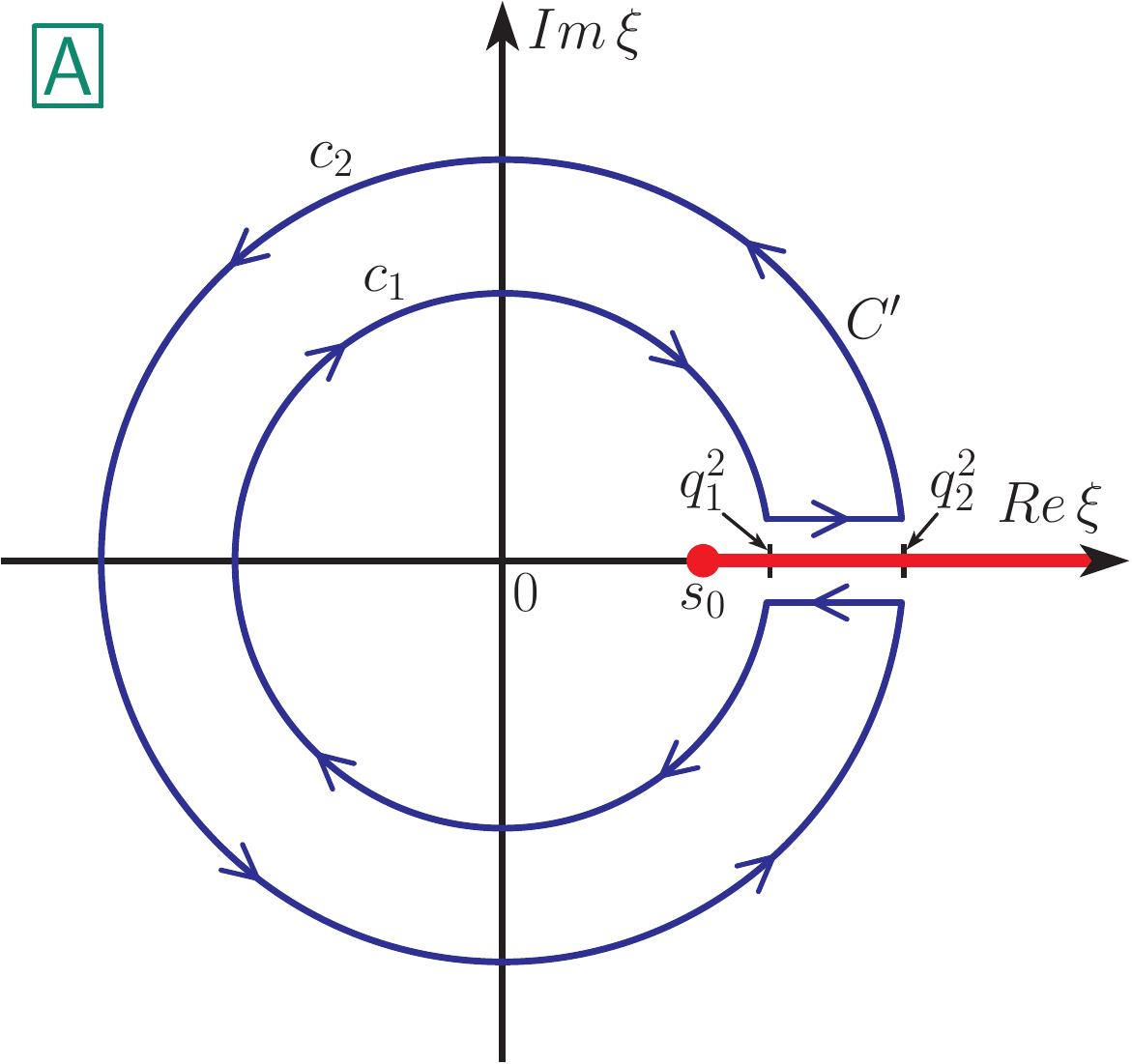}%
\hspace{10mm}%
\includegraphics[width=75mm,clip]{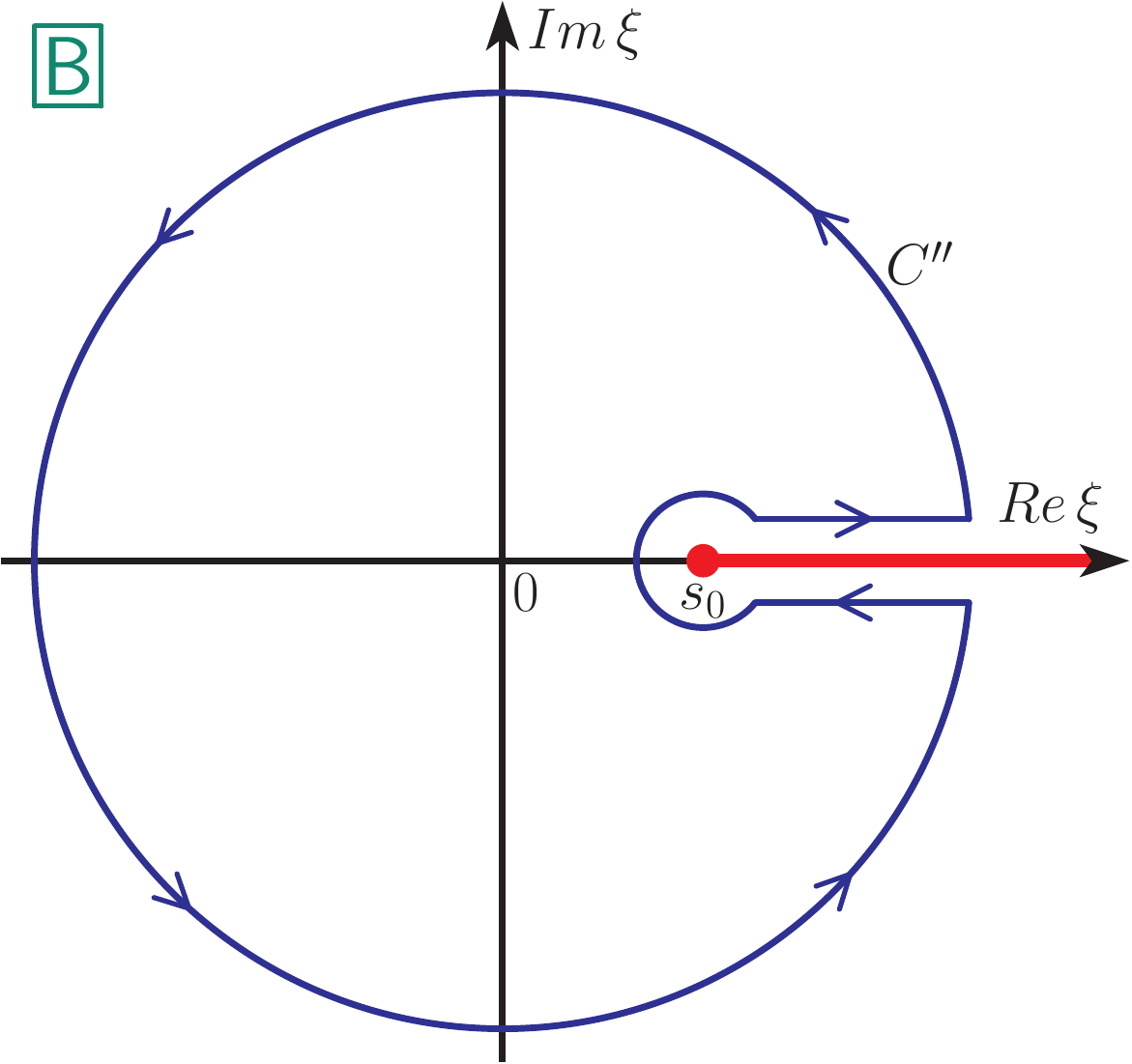}}
\caption{The closed integration contours in the complex~$\xi$--plane in
Eq.~(\ref{PDispExt}) in the presence (contour~$C'$, plot~``A'') and in the
absence (contour~$C''$, plot~``B'') of the quark flavour thresholds. The
physical cut~$\xi \ge s_{0}$ of the hadronic vacuum polarization
function~$\Pi(\xi)$~(\ref{P_Def}) is shown along the positive semiaxis of
real~$\xi$.}
\label{Plot:CGDRextPR}
\end{figure}

Two remaining dispersion relations~(\ref{PDisp}) and~(\ref{GDR_DR}), being
commonly employed for the data--driven evaluation of the functions~$\Pi(q^2)$
and~$D(Q^2)$, are of a particular interest. To~derive the dispersion relation
for the hadronic vacuum polarization function in the presence of the quark
flavour thresholds it is convenient to use the once--subtracted Cauchy's
integral formula
\begin{equation}
\label{PDispExt}
\Pi(q^2) - \Pi(q_0^2) =
\frac{1}{2\pi i}(q^{2}-q_{0}^{2})\!\oint_{C'}\!
\frac{\Pi(\xi)}{(\xi-q^2)(\xi-q_0^2)}\, d\xi,
\end{equation}
where the closed integration contour~$C'$ in the complex $\xi$--plane is
shown in Fig.~\ref{Plot:CGDRextPR}$\,$A. This equation can be represented~as
(the limit $\varepsilon \to 0_{+}$ is assumed in what follows)
\begin{align}
\label{GDRextPRint1}
\Pi(q^2) - \Pi(q_0^2) & =
\frac{q^{2}-q_{0}^{2}}{2\pi i}
\!\int\limits_{q_{1}^{2}+\iep}^{q_{2}^{2}+\iep}\!
\frac{\Pi(\xi)}{(\xi-q^2)(\xi-q_0^2)}\, d\xi +
\frac{q^{2}-q_{0}^{2}}{2\pi i}
\!\int_{\!c_{2}}\! \frac{\Pi(\xi)}{(\xi-q^2)(\xi-q_0^2)}\, d\xi
\nonumber \\[1mm]
& +
\frac{q^{2}-q_{0}^{2}}{2\pi i}
\!\int\limits_{q_{2}^{2}-\iep}^{q_{1}^{2}-\iep}\!
\frac{\Pi(\xi)}{(\xi-q^2)(\xi-q_0^2)}\, d\xi +
\frac{q^{2}-q_{0}^{2}}{2\pi i}
\!\int_{\!c_{1}}\! \frac{\Pi(\xi)}{(\xi-q^2)(\xi-q_0^2)}\, d\xi.
\end{align}
Then, the change of the integration variables~$\xi = \sigma + \iep$ and~$\xi
= \sigma - \iep$ in, respectively, first and third terms on the right--hand
side of Eq.~(\ref{GDRextPRint1}) casts them~to
\begin{equation}
\frac{q^{2}-q_{0}^{2}}{2\pi i}
\!\int\limits_{q_{1}^{2}+\iep}^{q_{2}^{2}+\iep}\!
\frac{\Pi(\xi)}{(\xi-q^2)(\xi-q_0^2)}\, d\xi =
\frac{q^{2}-q_{0}^{2}}{2\pi i}
\!\int\limits_{q_{1}^{2}}^{q_{2}^{2}}\!
\frac{\Pi(\sigma+\iep)}{(\sigma-q^{2})(\sigma-q_{0}^{2})}\, d\sigma
\end{equation}
and
\begin{equation}
\frac{q^{2}-q_{0}^{2}}{2\pi i}
\!\int\limits_{q_{2}^{2}-\iep}^{q_{1}^{2}-\iep}\!
\frac{\Pi(\xi)}{(\xi-q^2)(\xi-q_0^2)}\, d\xi =
-\frac{q^{2}-q_{0}^{2}}{2\pi i}
\!\int\limits_{q_{1}^{2}}^{q_{2}^{2}}\!
\frac{\Pi(\sigma-\iep)}{(\sigma-q^{2})(\sigma-q_{0}^{2})}\, d\sigma.
\end{equation}
In~turn, the change of the integration variables~$\xi =
q_{2}^{2}e^{i\varphi}$ and~$\xi = q_{1}^{2}e^{i\varphi}$ in, respectively,
second and fourth terms on the right--hand side of Eq.~(\ref{GDRextPRint1})
leads~to
\begin{equation}
\frac{q^{2}-q_{0}^{2}}{2\pi i}
\!\int_{\!c_{2}}\! \frac{\Pi(\xi)}{(\xi-q^2)(\xi-q_0^2)}\, d\xi =
J(q^{2}, q_{0}^{2}, q_{2}^{2})
\end{equation}
and
\begin{equation}
\frac{q^{2}-q_{0}^{2}}{2\pi i}
\!\int_{\!c_{1}}\! \frac{\Pi(\xi)}{(\xi-q^2)(\xi-q_0^2)}\, d\xi =
- J(q^{2}, q_{0}^{2}, q_{1}^{2}),
\end{equation}
where
\begin{equation}
\label{Jdef}
J(q^{2}, q_{0}^{2}, p^{2}) = \frac{q^{2}-q_{0}^{2}}{2\pi}
\!\int\limits_{\varepsilon}^{2\pi-\varepsilon}\!
\frac{\Pi(p^{2}e^{i\varphi})p^{2}e^{i\varphi}}
{(p^{2}e^{i\varphi}-q^{2})(p^{2}e^{i\varphi}-q_{0}^{2})}\, d\varphi.
\end{equation}
Therefore, Eq.~(\ref{GDRextPRint1}) acquires the following~form
\begin{equation}
\label{GDRextPRint2}
\Pi(q^2) - \Pi(q_0^2) = (q^{2}-q_{0}^{2})
\!\int\limits_{q_{1}^{2}}^{q_{2}^{2}}\!
\frac{R(\sigma)}{(\sigma-q^{2})(\sigma-q_{0}^{2})}\, d\sigma +
J(q^{2}, q_{0}^{2}, q_{2}^{2}) - J(q^{2}, q_{0}^{2}, q_{1}^{2}),
\end{equation}
with Eq.~(\ref{GDRextRP}) being employed.

In~the absence of the quark flavour thresholds (i.e.,~for the continuous
hadronic vacuum polarization function in the whole energy range) the
integration contour in~Eq.~(\ref{PDispExt}) can be chosen in the form
displayed in Fig.~\ref{Plot:CGDRextPR}$\,$B, that eventually results in the
dispersion relation~(\ref{PDisp}). Specifically, in this case the integration
along the edges of the cut yields the right--hand side of Eq.~(\ref{PDisp}),
whereas the contributions of the integrals along the circles of infinitely
large and infinitely small radii vanish. At~the same time, in the presence of
the quark flavour thresholds (i.e.,~for the piecewise continuous hadronic
vacuum polarization function) the~dispersion relation receives additional
contributions~(\ref{GDRextPRint2}), namely
\begin{equation}
\label{GDRextPR}
\Pi(q^2) - \Pi(q_0^2) =
(q^2-q_{0}^{2})\!\int\limits_{s_{0}}^{\infty}\!
\frac{R(\sigma)}{(\sigma-q^2)(\sigma-q_0^2)}\, d\sigma +
\sum\limits_{f} \Delta J_{[f]}(q^{2}, q_{0}^{2}, q_{f}^{2}).
\end{equation}
In~this equation the sum runs over the quark flavour thresholds,
\begin{equation}
\label{DltJdef}
\Delta J_{[f]}(q^{2}, q_{0}^{2}, q_{f}^{2}) =
J_{[f-1]}(q^{2}, q_{0}^{2}, q_{f}^{2}) -
J_{[f]}(q^{2}, q_{0}^{2}, q_{f}^{2}),
\end{equation}
$q_{f}^{2}$~denotes the squared threshold mass, and the subscript~``$[f]$''
indicates that in Eq.~(\ref{Jdef}) the hadronic vacuum polarization
function~$\Pi(q^2)$ is calculated for $\nf=f$ active flavours.

In~turn, the corresponding relation for the Adler function~$D(Q^2)$ can be
directly obtained from Eqs.~(\ref{GDRextDP}) and~(\ref{GDRextPRint2}),
specifically
\begin{equation}
\label{GDRextDRint}
D(Q^2) = Q^2 \!\int\limits_{q_{1}^{2}}^{q_{2}^{2}}\!
\frac{R(\sigma)}{(\sigma+Q^2)^2}\, d\sigma +
K(Q^2,q_{2}^{2}) - K(Q^2,q_{1}^{2}),
\end{equation}
where
\begin{equation}
\label{Kdef}
K(Q^{2}, p^{2}) = \frac{Q^{2}}{2\pi}
\!\int\limits_{\varepsilon}^{2\pi-\varepsilon}\!
\frac{\Pi(p^{2}e^{i\varphi})p^{2}e^{i\varphi}}
{(p^{2}e^{i\varphi}+Q^{2})^{2}}\, d\varphi.
\end{equation}
In~the absence of the quark flavour thresholds (i.e.,~for the continuous
hadronic vacuum polarization function in the entire energy range) the
integration limits in Eq.~(\ref{GDRextDRint}) can safely be set to~$q_{1}^{2}
\to s_{0}$ and~$q_{2}^{2} \to \infty$, that leads to the dispersion
relation~(\ref{GDR_DR}). However, the presence of the quark flavour
thresholds (i.e.,~the piecewise continuous hadronic vacuum polarization
function) ultimately generates additional contribution, namely
\begin{equation}
\label{GDRextDR}
D(Q^2) = Q^2 \!\int\limits_{s_{0}}^{\infty}\!
\frac{R(\sigma)}{(\sigma+Q^2)^2}\, d\sigma +
\sum\limits_{f} \Delta K_{[f]}(Q^2, q_{f}^{2}).
\end{equation}
In~this equation the sum runs over the quark flavour thresholds,
\begin{equation}
\label{DltKdef}
\Delta K_{[f]}(Q^2, q_{f}^{2}) =
K_{[f-1]}(Q^2, q_{f}^{2}) - K_{[f]}(Q^2, q_{f}^{2}),
\end{equation}
$q_{f}^{2}$~stands for the squared threshold mass, and the
subscript~``$[f]$'' indicates that in Eq.~(\ref{Kdef}) the hadronic vacuum
polarization function~$\Pi(q^2)$ is calculated for $\nf=f$ active flavours.

\section{Conclusions}
\label{Sect:Concl}

The equivalent representations for the hadronic vacuum polarization
contributions to the muon anomalous magnetic moment~$a^{\text{HVP}}_{\mu}$ in
the presence of the quark flavour thresholds are studied. Specifically, the
explicit relations between the contributions given by the integration over a
finite kinematic interval to~$a^{\text{HVP}}_{\mu}$ expressed in terms of the
hadronic vacuum polarization function, Adler function, and
the~\mbox{$R$--ratio} of electron--positron annihilation into hadrons are
derived [Eqs.~(\ref{RelPRint2}) and~(\ref{RelPDint})]. It is shown that the
quark flavour thresholds of the hadronic vacuum polarization function
generate additional contributions to~$a^{\text{HVP}}_{\mu}$ expressed in
terms of the Adler function and the~\mbox{$R$--ratio} and the explicit
expressions for such contributions are obtained [Eqs.~(\ref{AmuPR})
and~(\ref{AmuPD})]. The~commonly employed dispersion relations, which bind
together hadronic vacuum polarization function, Adler function,
and~\mbox{$R$--ratio}, are extended to account for the effects of the quark
flavour thresholds (Sect.~\ref{Sect:GDRext}). The~examined additional
contributions due to the heavy quark thresholds to~$a^{\text{HVP}}_{\mu}$
expressed in terms of the~\mbox{$R$--ratio} appear to be quite sizable, that
can be of a particular relevance for the data--driven approach to the
assessment of the hadronic part of the muon anomalous magnetic moment.

\end{document}